\documentclass[letterpaper,11pt]{article}
\pdfoutput=1
\usepackage{jheppub}
\usepackage{slashed}
\usepackage{graphicx} 
\usepackage{amsmath} 
\usepackage{multirow} 
\usepackage{hyperref}
\usepackage{epstopdf}
\usepackage{mathtools}

\newcommand{\beq}{\begin{equation}}
\newcommand{\eeq}{\end{equation}} 
\newcommand{\bea}{\begin{eqnarray}}
\newcommand{\eea}{\end{eqnarray}}
\newcommand{\ba}{\begin{array}}
\newcommand{\ea}{\end{array}}

\def\m1{M_1}
\def\m2{M_2}
\def\m3{M_3}

\def\ch10{\tilde \chi^0_1}

\def\BR{\rm Br}

\def\gev{\,{\rm GeV}}

\def\Re{\,{\rm Re}}
\def\to{\rightarrow}

\newcommand{\lsim}{\mathrel{\mathop{\kern 0pt \rlap
  {\raise.2ex\hbox{$<$}}}
  \lower.9ex\hbox{\kern-.190em $\sim$}}} 
\newcommand{\gsim}{\mathrel{\mathop{\kern 0pt \rlap
  {\raise.2ex\hbox{$>$}}}
  \lower.9ex\hbox{\kern-.190em $\sim$}}}

\definecolor{pink}{RGB}{255,105,180}

\def\abi{\,{\rm ab}^{-1}}

\title{Probing the electroweak phase transition\\
 via enhanced di-Higgs boson production}

\author[a,b,c]{Marcela Carena}
\author[a]{, Zhen Liu}
\author[d,e]{and Marc Riembau}

\affiliation[a]{Theoretical Physics Department, Fermi National Accelerator Laboratory, Batavia, Illinois, 60510, USA}
\affiliation[b]{Enrico Fermi Institute, University of Chicago, Chicago, Illinois, 60637, USA}
\affiliation[c]{Kavli Institute for Cosmological Physics, University of Chicago, Chicago, Illinois, 60637, USA}
\affiliation[d]{IFAE and BIST, Universitat Aut\`onoma de Barcelona, E-08193~Bellaterra,~Barcelona, Spain}
\affiliation[e]{DESY, Notkestra{\ss}e 85, D-22607 Hamburg, Germany}
\emailAdd{carena@fnal.gov}
\emailAdd{zliu2@fnal.gov}
\emailAdd{marc.riembau@desy.de}

\abstract{
We consider a singlet extension of the Standard Model (SM) with a spontaneous $Z_2$ breaking and study 
the gluon-gluon fusion production of the heavy scalar, with subsequent decay into a pair of SM-like Higgs bosons. We find that an on-shell interference effect can notably enhance the resonant di-Higgs production rate up to 40\%. 
In addition, consistently taking into account both the on-shell and off-shell interference effects between the heavy scalar and the SM di-Higgs diagrams significantly improves the HL-LHC and HE-LHC reach in this channel. As an example, within an effective field theory analysis in an explicitly $Z_2$ breaking scenario, we further discuss the potential to  probe the parameter region compatible with a first order electroweak phase transition. 
Our analysis is applicable for general potentials of the singlet extension of the SM as well as for more general resonance searches.
}

\preprint{
\begin{flushright}
FERMILAB-PUB-17-600-T
\end{flushright}
}

\begin{document}
\maketitle
\flushbottom 

\section{INTRODUCTION}
Probing the intriguing possibility of electroweak baryogenesis~\cite{Klinkhamer:1984di,Kuzmin:1985mm,Arnold:1987mh,Arnold:1987zg,Khlebnikov:1988sr} becomes of higher relevance after the SM Higgs boson discovery at the LHC~\cite{Aad:2012tfa,Chatrchyan:2012xdj}. In such  mechanism, a strongly first-order electroweak phase transition (EWPT) is a crucial ingredient to maintain the matter-antimatter asymmetry generated at the electroweak scale~\cite{Sakharov:1967dj}. The SM Higgs potential is insufficient to provide such condition and many extensions of the SM have hence been proposed~\cite{Huet:1995sh,Carena:1996wj,Laine:1998qk,Laine:2000xu,Espinosa:1996qw,Carena:1997ki,Huber:2001xf,Cline:2000kb,Carena:2002ss,Lee:2004we,Cirigliano:2009yd,Carena:2008rt,Quiros:1999tx,Delepine:1996vn,Wainwright:2012zn,Quiros:1999jp,Kotwal:2016tex,Curtin:2014jma,Huang:2015tdv,Huang:2017jws,Chen:2017qcz,Goncalves:2017iub,Huang:2016cjm,Curtin:2016urg,Jiang:2015cwa,Cheng:2018ajh,Fuyuto:2014yia,Basler:2016obg,Basler:2017uxn,Goertz:2013kp,Goertz:2014qta,Huang:2015izx,Goertz:2015dba,DiLuzio:2017tfn}.
%,Huang:2016odd,Cao:2017oez}.
Generically, including additional bosonic degrees of freedom with sizable coupling strength to the SM Higgs boson can increase the barrier between the broken and unbroken electroweak vacua at the critical temperature of the phase transition, see e.g., Ref.~\cite{Katz:2014bha}. Amongst many of the possibilities, the singlet scalar extension  of the SM is of particular interest~\cite{Curtin:2014jma,Huang:2015tdv,Huang:2017jws,Chen:2017qcz}. Due to its singlet nature, the scalar is hard to be probed at the  LHC.  Therefore the singlet SM extension  serves as the simplest, yet elusive benchmark to test  a sufficiently strong  first-order phase transition compatible with the Higgs boson mass measurements at the LHC~\cite{Curtin:2014jma,Kotwal:2016tex}. 

The scalar potential of  a real singlet scalar extension of the SM can be further categorized into three types, depending on  the behavior of the real singlet $s$ under the $Z_2$ parity operation $s\rightarrow -s$, namely: the $Z_2$ symmetric, the spontaneous $Z_2$ breaking, and the general potential. The $Z_2$ symmetric potential leads to a stable singlet scalar, resulting in the singlet being a possible dark matter candidate and yielding missing energy signals at colliders~\cite{Curtin:2014jma}.
Without $Z_2$ protection, the singlet would mix with the SM Higgs and (in most cases) a promptly decaying scalar particle would provide a  rich phenomenology at colliders. The singlet scalar could be produced resonantly  and decay back to pairs of SM particles, dominantly into $WW$, $ZZ$, $HH$ and $t\bar t$. The signal of  a singlet scalar resonance decaying into $HH$ is a smoking-gun  for singlet enhanced EWPT~\cite{Chen:2014ask,Dawson:2015haa,Chen:2017qcz,Huang:2017jws,Robens:2016xkb,Lewis:2017dme,Dawson:2016ugw,Huang:2017nnw,deFlorian:2016spz}.

Searches for resonant di-Higgs production have received much attention by both the ATLAS and CMS collaborations~\cite{Aaboud:2016xco,Aad:2015xja,Sirunyan:2017djm,CMS-PAS-HIG-17-006,CMS-PAS-HIG-17-008,CMS-PAS-HIG-17-009}. In the case of a singlet resonance,  constraints from SM precision measurements  render these searches more challenging. From one side  precision measurements  imply that  the singlet-doublet mixing parameter is constrained to be small over a large region of parameter space.  From the other side, the singlet only couples to SM particles through mixing with the SM Higgs doublet. This results in a reduced di-Higgs production via singlet resonance decays. In particular, the singlet resonance amplitude  becomes of the same order as the SM  triangle  and box diagram amplitudes. Most important, in this work we shall show that a large relative phase between the SM box diagram and the singlet triangle diagram becomes important. This special on-shell interference  effect  has been  commonly overlooked in the literature and turns out to have important phenomenological implications.
We shall choose the spontaneous $Z_2$ breaking scenario of the SM plus singlet to demonstrate the importance of the novel on-shell interference effect for the resonant singlet scalar searches in the di-Higgs production mode.

Our paper is organized as follows. In Sec.~\ref{sec:model}, we study the details of the $Z_2$ symmetric potential with spontaneous $Z_2$ and electroweak symmetry breaking, including its parametric dependence on physical quantities,  requirements  of vacuum stability and perturbative unitarity, and the decay properties of the singlet scalar. In Sec.~\ref{sec:interference}, we present a detailed discussion on various types of interference effects for the process  $gg\to HH$. We focus on the special case of the on-shell interference effect for the resonant singlet production and  show their parametric dependence. In Sec.~\ref{sec:pheno}, we study the impact of the on-shell interference effect for the High Luminosity (HL)- and High Energy (HE)- LHC searches and address  how a comprehensive study of the on-shell and off-shell differential cross sections would improve on their sensitivity. This in turn will provide the tools for a much more robust test  of a strongly first-order  EWPT in this type of models. We reserve Sec.~\ref{sec:conclude} for our conclusions.

%Strongly first order electroweak phase transition...

%We consider a simplest extension of the SM with a spontaneous breaking $Z_2$ global symmetry amongst the scalar sectors.

\section{MODEL FRAMEWORK}
\label{sec:model}

We will consider the simplest extension of the SM that can assist the scalar potential to induce a strongly first-order electroweak phase transition, consisting of an additional real scalar singlet with a $Z_2$ symmetry. The scalar potential of the model can be written as
\beq
V(s,\phi)=-\mu^2 \phi^\dagger \phi -\frac 1 2 \mu_s^2 s^2 +  \lambda (\phi^\dagger \phi)^2 + \frac {\lambda_s} {4} s^4 + \frac {\lambda_{s\phi}} 2 s^2 \phi^\dagger\phi,
\label{eq:potential}
\eeq
where $\phi$ is the SM doublet
\footnote{ $\phi^T=(G^+,\frac 1 {\sqrt{2}} (h+ i G^0 +v))$, where $G^{\pm,0}$ are the Goldstone modes.}
and $s$ represents the new real singlet field. In the above, we adopt the conventional normalization for the couplings of the SM doublets and match the other couplings with the singlet with identical normalization. We allow for spontaneous $Z_2$ breaking with the singlet $s$ acquiring a vacuum expectation value (vev) $v_s$, since this case allows for interesting collider phenomenology of interference effects. As we shall show  later, the (on-shell) interference effects commonly exist for  loop-induced processes in BSM phenomenology and it is the focus of this paper. The CP even neutral component $h$ of the Higgs doublet field $\phi$ mixes with the real singlet scalar $s$, defining the new mass eigenstates $H$ and $S$
\bea
\binom{h}{s} =  \begin{pmatrix}
 \cos\theta & \sin\theta \\
 -\sin\theta & \cos\theta 
 \end{pmatrix}
 \binom{H}{S},
\eea
where $\theta$ is the mixing angle between these fields.
% and $h$ does not include the Higgs vev of $v$. 
%The capitalized $H$ and $S$ represent mass eigenstates that are mostly from the SM doublet $\phi$ and mostly singlet $s$, respectively. 
The five free parameters in Eq.~(\ref{eq:potential}) can  be traded by the two boundary conditions 
\beq
m_{H}= 125~\gev,~~v=246~\gev
\eeq
and the three ``physical'' parameters,
\beq
m_S,~~\tan\beta(\equiv \frac  {v_s} v),~{\rm and~}\sin\theta,\label{eq:basis}
\eeq
where $\tan\beta$ characterizes the ratio between the vevs of the doublet and the singlet scalar fields, respectively.

As a result, the parameters in the scalar potential in Eq.~(\ref{eq:potential}) can be expressed as functions of these new parameters, 
\bea
\mu^2 &=& \frac 1 4 \left(2m_H^2 \cos^2\theta + 2m_S^2 \sin^2\theta + (m_S^2-m_H^2) \tan\beta \sin2\theta\right)\\
\mu_s^2 &=& \frac 1 4 \left(2m_H^2 \sin^2\theta + 2m_S^2 \cos^2\theta + (m_S^2-m_H^2) \cot\beta\sin2\theta \right)\\
\lambda &=& \frac {m_H^2 \cos^2\theta +m_S^2\sin^2\theta} {2 v^2}\\\label{eq:lambdas} 
\lambda_s &=& \frac {m_H^2 \sin^2\theta +m_S^2\cos^2\theta} {2 \tan^2\beta~v^2}\\
\lambda_{s\phi} &=& \frac {\left(m_S^2-m_H^2\right) \sin2\theta} {2\tan\beta~v^2}.
\eea
%The choices of the normalization factors in the original potential
%As we can see clearly from above expressions, a very symmetric pattern appears due to our choice of the normalization factors of the couplings in the scalar potential.
Observe that the condition of spontaneous symmetry breaking implies that dimensionful quantities $\mu^2$ and $\mu_s^2$ can be directly expressed in terms of the original quartic couplings and the vevs,
\beq
\mu^2 = v^2 \left(\lambda + \frac 1 2\tan^2\beta \lambda_{s\phi}\right)\,,\quad\quad
\mu_s^2 = v^2\left(\tan^2\beta \lambda_s + \frac 1 2\lambda_{s\phi}\right).
\label{eq:muandmus}
\eeq

\subsection{Stability, unitarity and EWSB conditions}
\label{sec:stabilityunitarityandEWSBconditions}

\begin{figure}[thpb]
  \centering
  \includegraphics[width=.48\textwidth]{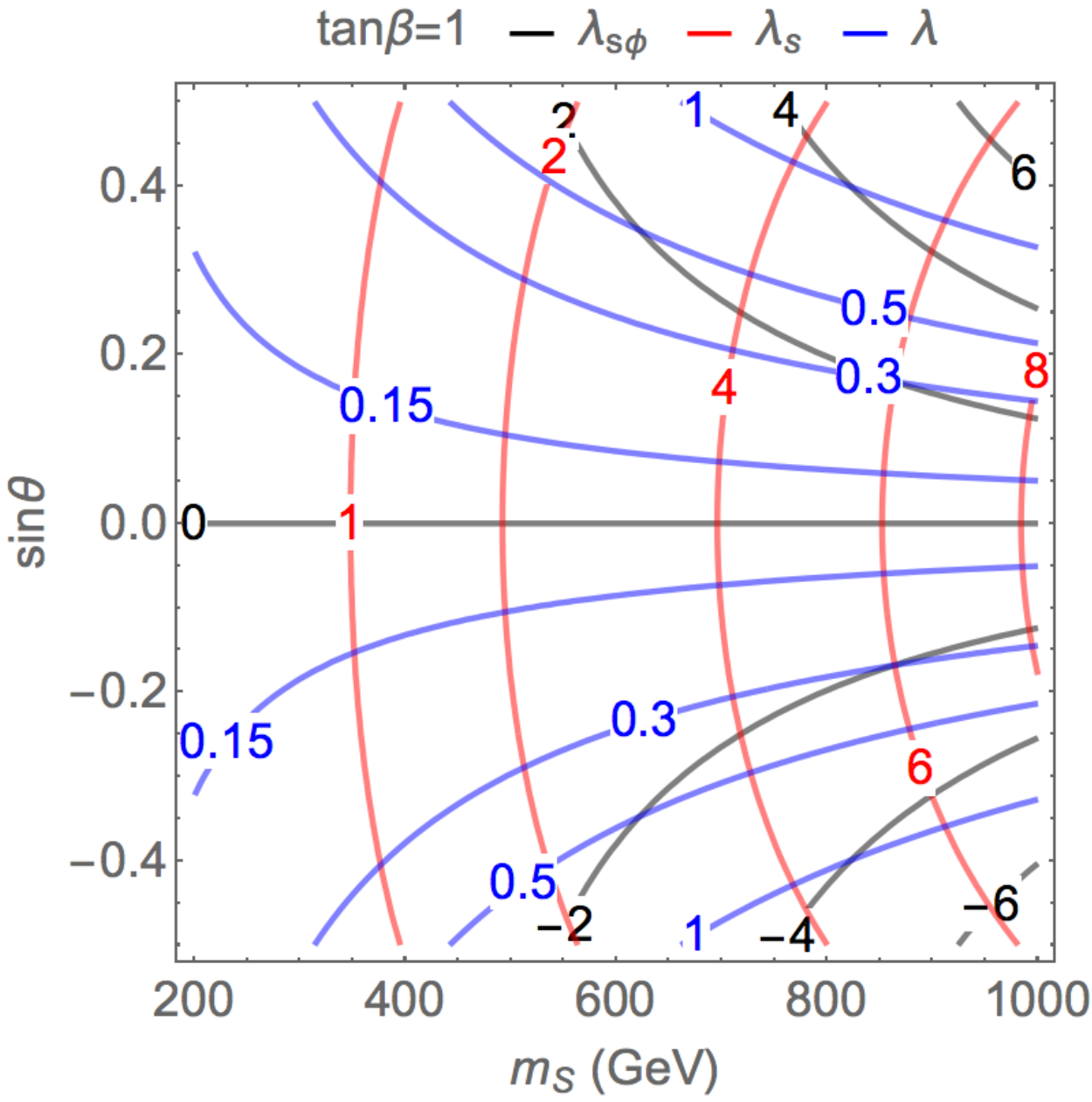} 
  \includegraphics[width=.48\textwidth]{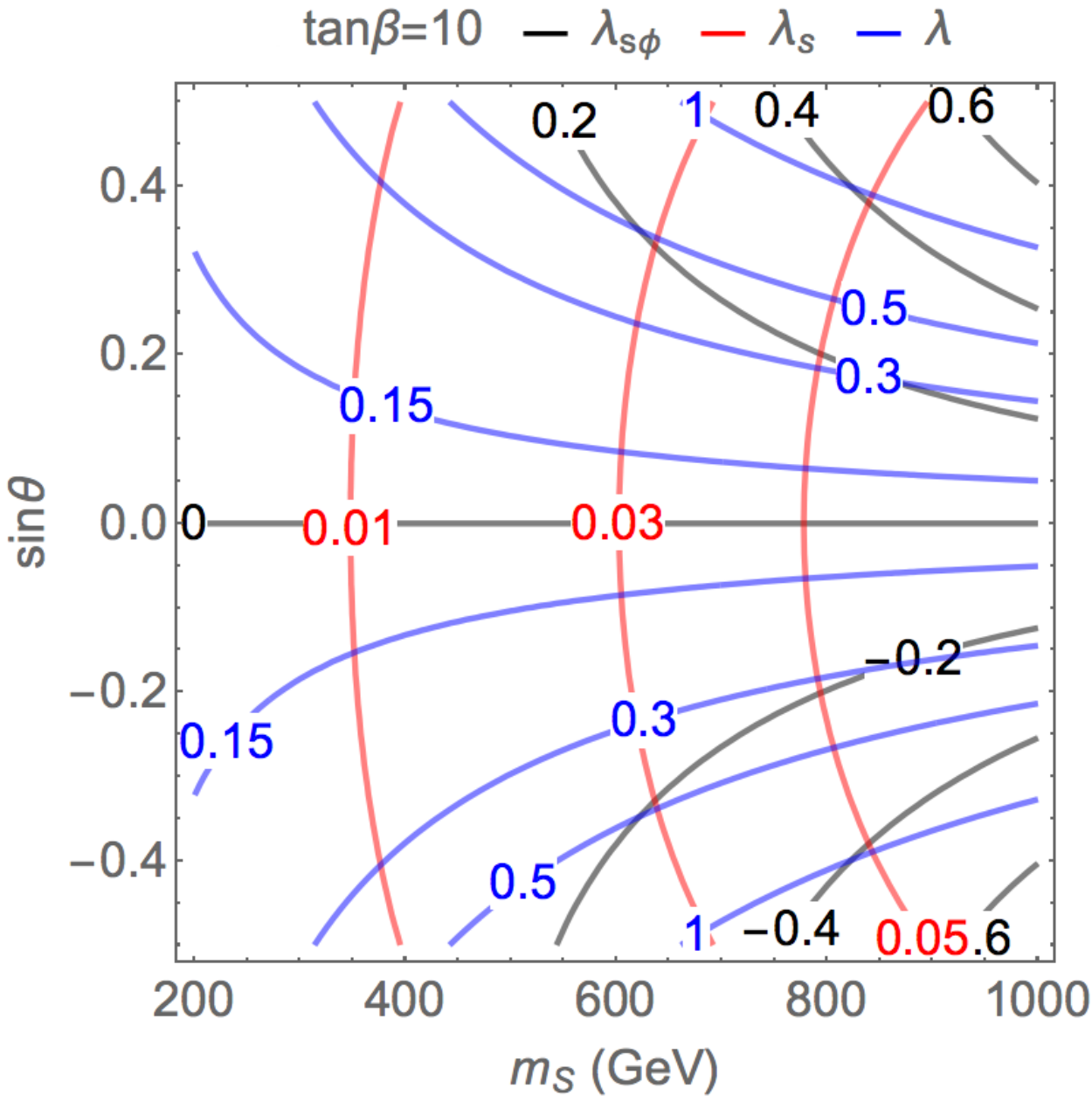}
  \caption{The values of the quartic couplings $\lambda$, $\lambda_s$ and $\lambda_{s\phi}$ as a function of the singlet-like scalar mass $m_S$ and the mixing angle $\sin\theta$ shown in blue, red and black contours, respectively. The left and right panels correspond to $\tan\beta$ values of 1 and 10, respectively. 
  %{\color{cyan} The shaded region are disallowed by vacuum stability and perturbative unitarity argument.}
  }
  \label{fig:lambda}
\end{figure}

It is useful to understand the quartic couplings in the potential in Eq.~(\ref{eq:potential}) in terms of  the physical parameters defined in Eq.~(\ref{eq:basis}), since the physical parameters make a straightforward connection with collider physics. %In \autoref{fig:lambda} we show the values of quartic couplings in the bare Lagrangian in \autoref{eq:potential} as a function of the heavy singlet-like scalar mass $m_S$ and the singlet-doublet mixing angle $\sin\theta$ for various values of $\tan\beta$. 
In Fig.~\ref{fig:lambda} we show the three independent quartic couplings in the $Z_2$-symmetric potential $\lambda$, $\lambda_s$ and $\lambda_{s\phi}$ in blue, red, and black contours, respectively, as a function of the heavy singlet-like scalar mass $m_S$ and the singlet-doublet mixing angle $\sin\theta$ for $\tan\beta$ = 1 (left panel) and  $\tan\beta$ = 10 (right panel). 
%We also show the limits from {\color{cyan} perturbative unitarity constraints on the quartics in gray shaded region, and} electroweak precision constraints in brown shaded region. 
As shown in the red contours, for low values of $\tan\beta$, a large quartic $\lambda_s$ is needed to obtain a heavy singlet, due to the fact that the singlet mass and its vev are related via its quartic coupling, see Eq.~(\ref{eq:lambdas}). However, the correlation between the singlet quartic, its mass and its vev  is only mildly dependent on the mixing angle $\sin\theta$. %The opposite is true 
A different behavior occurs for the Higgs quartic, in blue, being independent of the singlet vev but sensitive to the mixing angle.% and to the singlet mass for masses below 600 GeV
%As shown in these plots, the precision tests cover larger mixing angles of $\sin\theta>0.35$ for heavy singlet scalar mass above 350~GeV.\\

The stability of the potential and the perturbative unitarity arguments set constraints on the allowed sizes and signs of the quartic couplings that we will discuss now. The requirement of the potential being bounded from below leads to the conditions to the quartic couplings
\beq
\lambda,~\lambda_s > 0{\rm ~and~}\lambda_{s\phi} > -2\sqrt{\lambda \lambda_s}.
\label{eq:frombelow}
\eeq
The positivity of the Higgs and singlet quartic couplings is understood by considering large field values in the directions $\{ h, 0\}$ and $\{ 0, s\}$. The extra condition arises from considering large field values in an arbitrary direction. We see that negative values for the mixing quartic coupling $\lambda_{s\phi}$ are allowed if the other two quartics are large enough.

Furthermore, the spontaneous $Z_2$ and electroweak symmetry breaking vacuum $<\phi,s>=\{v/\sqrt{2}, v_s\}$ is a global minimum if the following is satisfied
\beq
\lambda_{s\phi} < +2\sqrt{\lambda \lambda_s}.
\label{eq:fromabove}
\eeq
For larger values of $\lambda_{s\phi}$, the electroweak and $Z_2$ breaking vacuum becomes a saddle point and the minima are located at $<\phi,s>=\{v/\sqrt{2}, 0\}$ and $<\phi,s>=\{ 0, v_s\}$ {\footnote {Note that the expressions for the vevs in terms of model parameters differ for the different extrema under discussion. We denote them using the same symbols, $v/\sqrt{2}$ and $v_s$,  since the discussion in this section does not rely on their precise values.}}. Observe that for positive values of $\mu^2$ and $\mu_s^2$ the origin $<\phi,s>=\{ 0,0 \}$ is always a maximum.

The conditions Eq.~(\ref{eq:frombelow}) and Eq.~(\ref{eq:fromabove}) have the following physical interpretation. The determinant of the mass matrix at the electroweak and $Z_2$ breaking minimum is proportional to $4\lambda \lambda_s-\lambda_{s\phi}^2$, which is equivalent to the previous requirements. When the determinant of the mass matrix becomes negative, and therefore one of the conditions fails, a tachyonic direction will be generated, destabilizing the system and evolving it to other minima.
From this perspective it is also clear that if we choose to work with $(v,m_H,m_S,\sin\theta,\tan\beta)$ as a set of parameters, the determinant of the mass matrix is just $m_H^2m_S^2$ and the requirements in this basis are automatically satisfied with physical masses. %Hence, there are no constraints coming from the stability of the potential. 
%Notice that for negative mixing angles one might get negative $\mu^2$ or $\mu_s^2$ (but not both) so the origin of the potential becomes a saddle point. However, if $4\lambda \lambda_s-\lambda_{s\phi}^2>0$ the electroweak or $Z_2$ breaking are induced by the mixing and $<\phi,s>=\{v/\sqrt{2}, v_s\}$ is still the global minimum.

\begin{figure}[t] 
	\centering
	\includegraphics[width=1\textwidth]{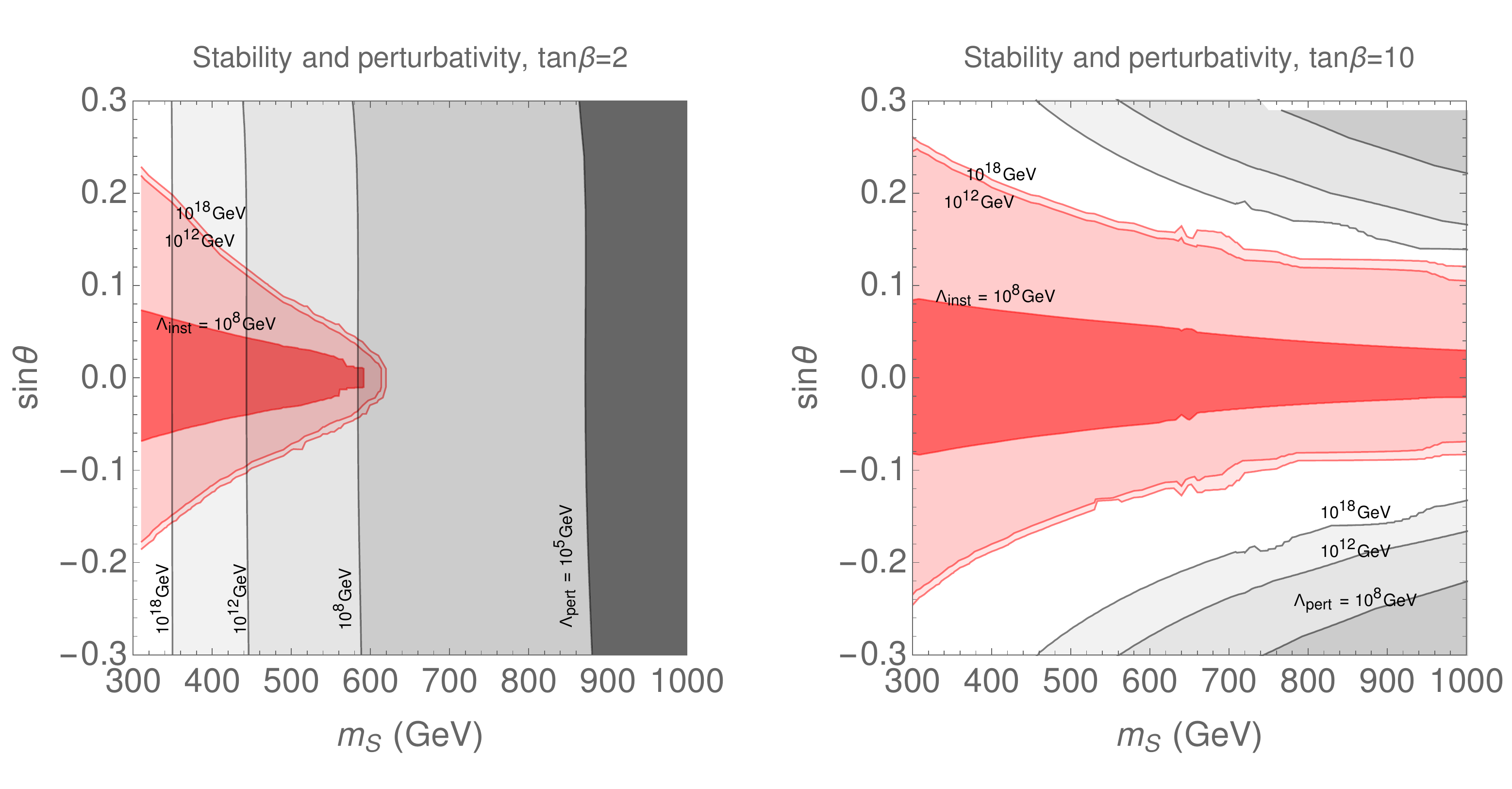}
  \caption{Regions of parameter space in the $m_S$-$\sin\theta$ plane disfavored by perturbativity and EW vacuum stability requirement at various scales. The gray shaded regions are disfavored by pertubative unitarity requirement at a given scale $\Lambda_{\rm pert}$. The red shaded regions correspond to regions disfavored by stability requirement at a given scale $\Lambda_{\rm inst}$. 
  %In the green region, the minimum is stable up to the Planck scale. For the point marked with a cross, we show in the right the evolution of the quartic couplings. The vertical line shows the scale where the quantity $4\lambda_\phi \lambda_s-\lambda_{s\phi}^2$ becomes negative.
  }
	\label{fig:thconstraints}
\end{figure}

The potential might be destabilized due to loop corrections, and the quantity $4\lambda \lambda_s-\lambda_{s\phi}^2$ might become negative at high scales. We study this effect taking into account the renormalization group evolution (RGE) of the quartic couplings given by the RGE equations in Sec.~\ref{sec:appthconstraints}. In Fig.~\ref{fig:thconstraints} we show, shaded in red, the region where the vacuum becomes unstable at a given energy scale.
%, once we include the RG evolution of the quartic couplings at one loop. %the consequences of including the RG evolution at one loop of the quartics. 
%We evolve the quartics and show the scale where the quantity $4\lambda \lambda_s-\lambda_{s\phi}^2$ becomes negative in units of GeV if we start to run the couplings at 1TeV. 
For small singlet masses and mixing angles the instability scale is not modified with respect to the SM (which is around $\sim 10^8$ GeV at one loop and relaxes to about $10^{11}$ GeV once two-loop RGE is included~\cite{Degrassi:2012ry}), but for larger mixings the singlet shifts the values of the couplings, pushing the scale of instability to larger values. For larger $\tan\beta$, as shown in the right panel  of Fig.~\ref{fig:thconstraints}, the singlet quartic coupling are smaller and the instability condition extends to a  larger region since the effect of the singlet is  insufficient to compensate the effect of the destabilizing  top Yukawa coupling.

There is also a constraint on the size of the quartic couplings  given by the perturbative unitarity arguments. The $2\to 2$ amplitudes $\mathcal{A}$ should satisfy
\beq
\frac{1}{16\pi s}\int_s^0 dt\,|\mathcal{A}| \,<\, \frac{1}{2}.
\label{eq:pertbound}
\eeq
This comes from decomposing the amplitude in partial waves and requiring it being consistent with the optical theorem. We consider the different scattering amplitudes among the components of the Higgs doublet and the singlet, and look for the combination of states giving the largest contribution to Eq.~(\ref{eq:pertbound}). This is done by  building a matrix that contains all the $2\to 2$ amplitudes among those states and taking the largest eigenvalue. We derive the constraints on the sizes of the quartic couplings by this method, giving further details  in Sec.~\ref{sec:appthconstraints}.

In Fig.~\ref{fig:thconstraints}, the  gray shaded regions show  the constraints from the perturbative unitarity arguments after including the RGE effects,  labeled by  the scale at which unitarity is broken. We observe that smaller values of $\tan\beta$ and larger singlet masses have a lower unitarity breaking scale.  This is  due to the fact  that larger singlet masses require larger singlet quartic couplings. In addition,  larger $\tan\beta$ corresponds to larger vevs of the singlet and yields larger masses  for smaller values of the  quartic couplings. Hence   perturbative unitarity arguments  are relaxed as $\tan\beta$ increases as well as for smaller values of $m_S$.%As a result, the constraints from perturbativity requirement are milder.

%\clearpage

\subsection{Properties of the singlet-like scalar}

In addition to the effect of  singlet-doublet mixing governed by $\sin \theta$, the relevant phenomenology of the production of di-Higgs final states is further characterized by two trilinear coupling parameters
\beq
\mathcal{L}\supset \lambda_{HHH} H^3 + \lambda_{SHH} S H^2.
\eeq
The dimensionful parameter $\lambda_{HHH}$ is the modified trilinear Higgs coupling and $\lambda_{SHH}$ is the heavy Scalar-Higgs-Higgs coupling that drives the heavy scalar $S$ decay into the di-Higgs  final state. Both couplings  can be written in terms of the physical parameters $m_s$, $\sin\theta$ and $\tan\beta$ as
\bea
\lambda_{HHH}&=& -\frac {m_H^2} {2\tan\beta~v} \left(\tan\beta \cos^3\theta-\sin^3\theta\right),\label{eq:111}\\
\lambda_{SHH}&=& -\frac {m_H^2} {2\tan\beta~v} \sin2\theta (\tan\beta~\cos\theta+ \sin\theta)(1+\frac {m_S^2}{ 2 m_H^2}).
\label{eq:211}
\eea

\begin{figure}[thpb]
  \centering
  \includegraphics[width=.48\textwidth]{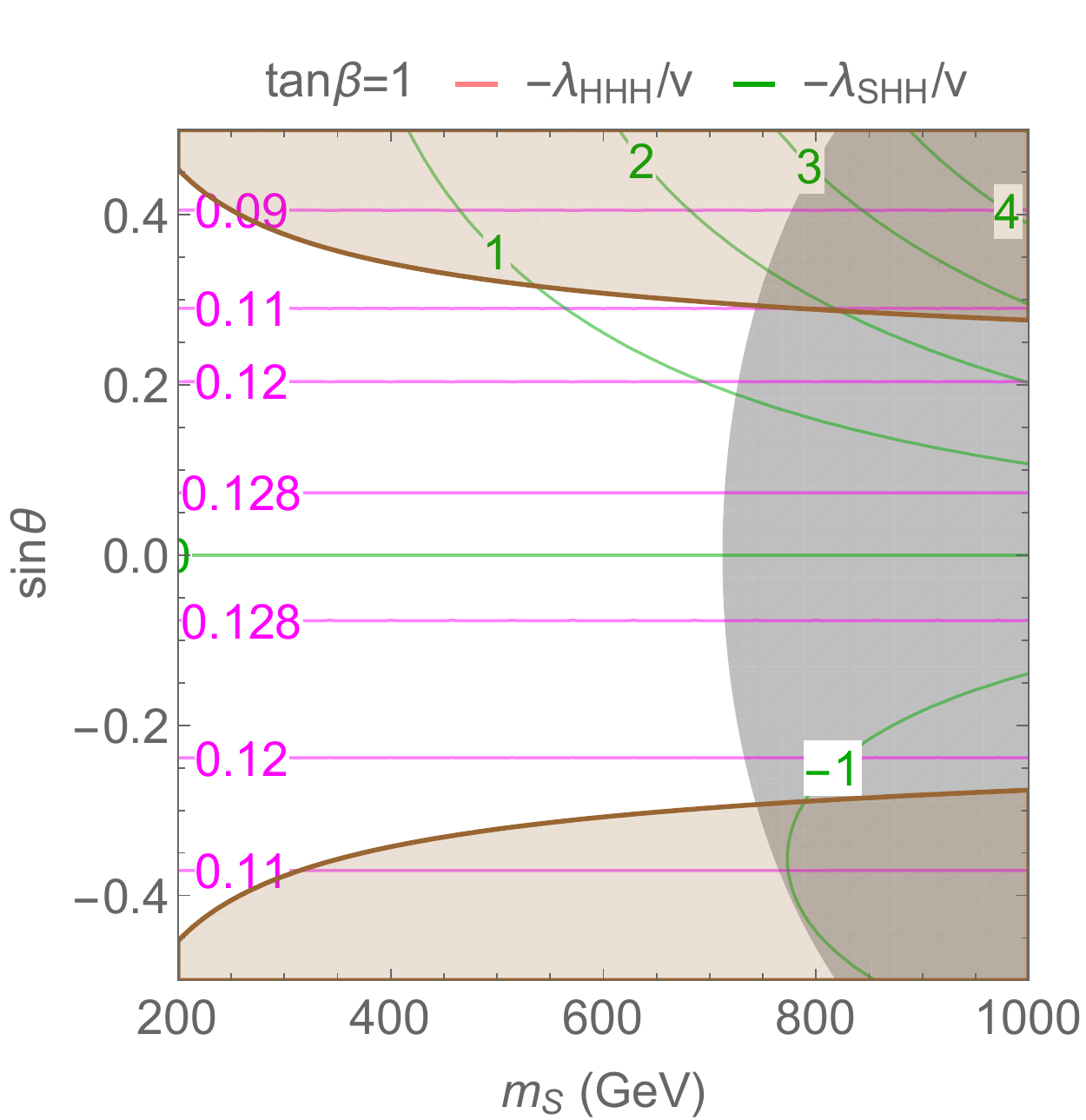}
  \includegraphics[width=.48\textwidth]{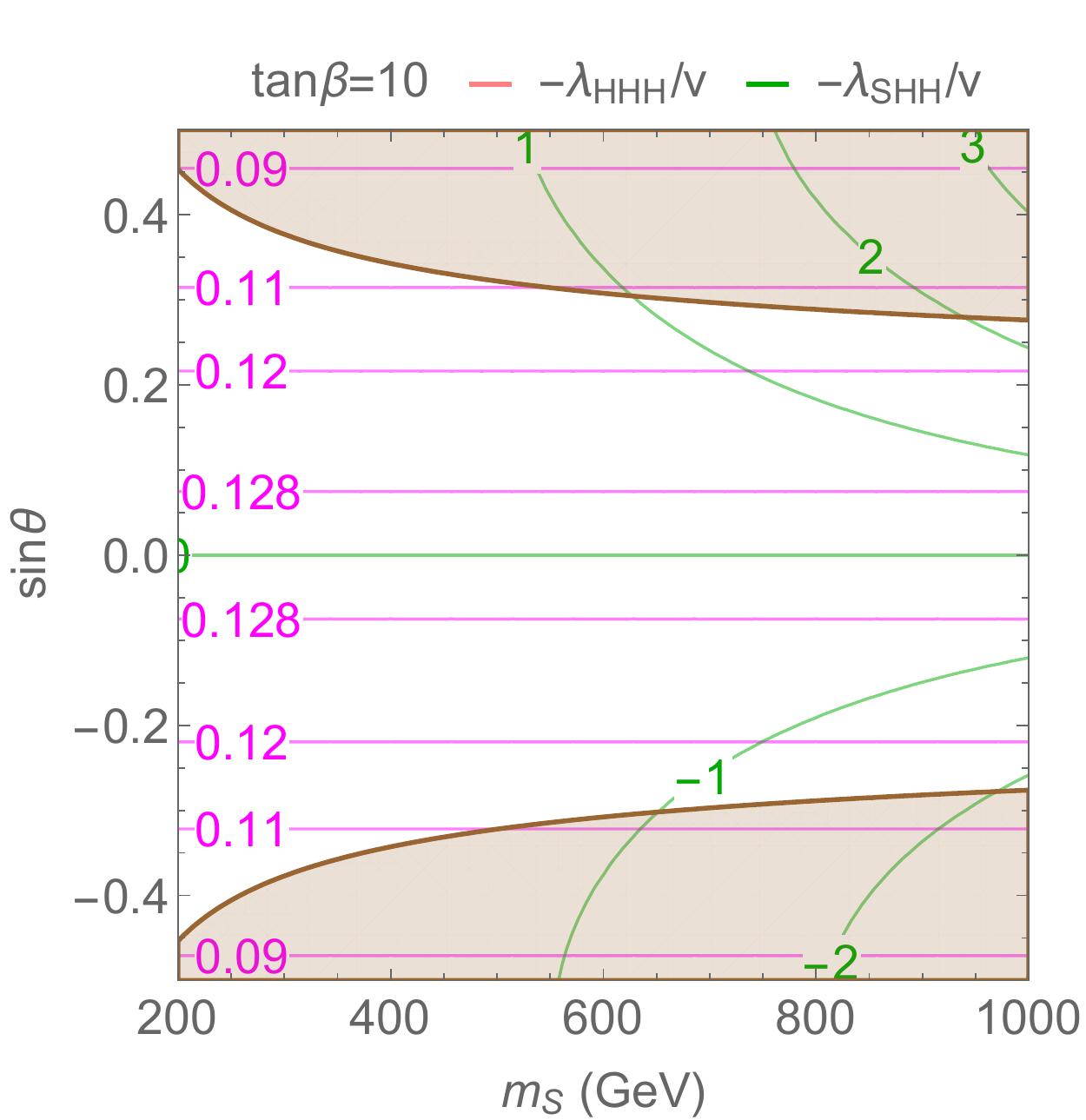}
  \caption{The phenomenlogically interesting trilinear scalar couplings, normalized by the SM doublet vev $v$, $-\lambda_{HHH}/v$ and $-\lambda_{SHH}/v$ as a function of the singlet-like scalar mass $m_S$ and the mixing angle $\sin\theta$ shown in magenta and dark green contours, respectively. The left and right panels correspond to $\tan\beta$ = 1 and 10, respectively. The gray shaded region is disallowed by vacuum stability and perturbative unitarity arguments, while the brown shaded regions are disallowed by EWPO.} 
  \label{fig:trilinear}
\end{figure}

In Fig.~\ref{fig:trilinear}, we show the values of trilinear couplings between mass eigenstates, $-\lambda_{HHH}/v$ and $-\lambda_{SHH}/v$ in green and magenta curves,
%after the $Z_2$- and EW-symmetry breaking in \autoref{eq:potential} 
as a function of the heavy singlet-like scalar mass $m_S$ and the singlet-doublet mixing angle $\sin\theta$ for $\tan\beta$ = 1 (left panel) and $\tan\beta$ = 10 (right panel). We can observe that the trilinear coupling of the SM-like Higgs remains insensitive to the singlet mass and receives moderate modifications with respect to its SM value. On the other hand, the trilinear $\lambda_{SHH}$ that determines the rate of the heavy scalar decay into Higgs pairs is quite sensitive to the precise value of the singlet-like scalar mass and the mixing angle $\sin\theta$.%, but is mostly insensitive to $\tan\beta$.

%At energies below the singlet mass, 
The heavy singlet mixing with the SM Higgs will induce a global shift on all the SM-like Higgs couplings. While this mixing does not change the SM branching ratios, the production rates of the Higgs boson will be reduced by a factor $\cos^2\theta$. %\ZL{Please rewrite the following sentences} 
The Higgs boson data from LHC at 7 and 8 TeV sets a constraint of $|\sin\theta|<0.36$ at 95\% C.L., independently of the singlet mass. %, while the projections on measurements of inclusive rates at HL-LHC will be able to set a constraint on $|\sin\theta|<0.25$ at 95\%CL.
The HL-LHC projection increases this limit very mildly due to the dominant effect from systematic and theory uncertainties. In addition, the current limit is driven by a measured  $\sim1$-$\sigma$ excess of signal strength over the SM Higgs expectation.
%The mild increase at the HL-LHC phase is because the relevant channels are dominated by systematic and theory uncertainties. 
Moreover, the singlet mixing affects the electroweak precision observables (EWPO) measured at LEP, setting slightly stronger limits than those coming from Higgs physics. Hence, in Fig.~\ref{fig:trilinear} we only show as brown shaded regions those excluded by EWPO, and refer to Sec.~\ref{sec:appthconstraints_indirect} for a detailed discussion. We also show in gray the region disallowed by vacuum stability and perturbative unitarity arguments at the scale $m_S$ where the physical parameters are defined, as discussed in the previous section.

Let's now discuss the decay properties of the singlet like scalar.  Its  decay to Higgs pairs is governed by  the trilinear coupling $\lambda_{SHH}$
\beq
\Gamma_S (S\to HH) = \frac {\lambda_{SHH}^2} {32 \pi m_S} \sqrt{ 1- \frac {4 m_H^2} {m_S^2}}
\label{eq:gammaS}
\eeq 
and to other SM particles via its mixing with the SM Higgs. The total singlet like scalar width can be written as,
\beq
\Gamma_S^{\rm tot} = \Gamma_S (S\to HH) + \sin^{2}\theta\ \Gamma_H^{\rm tot}|_{m_H\to m_S},
\eeq
where $\Gamma_H^{\rm tot}|_{m_H\to m_S}$ is the total width of a SM Higgs with  mass  $m_S$.

%It is important to understand the scaling behaviour of the partial width and the branching fractions to Higgs pairs for the heavy scalar state. From Eq.~\ref{eq:gammaS} and Eq.~\ref{eq:211}, we can see in general width scales quadratically in $\lambda_{SHH}$ and the coupling grows also quadratically in $m_S$ for large $m_S$. Hence in large scalar mass $m_S$, the particle width $\Gamma_S(S\to HH)$ scales as the third power of the scalar mass. 
%The partial width to $WW$, $ZZ$ through the mixing with the SM Higgs also grows as the third power of the scalar mass, due to the longitudinal enhancement for the massive vector gauge bosons. Consequently, the singlet branching fraction to $HH$ remain in the 20\%-40\% range over a large span of the parameter space. 

\begin{figure}[t]
  \centering
  \includegraphics[width=.48\textwidth]{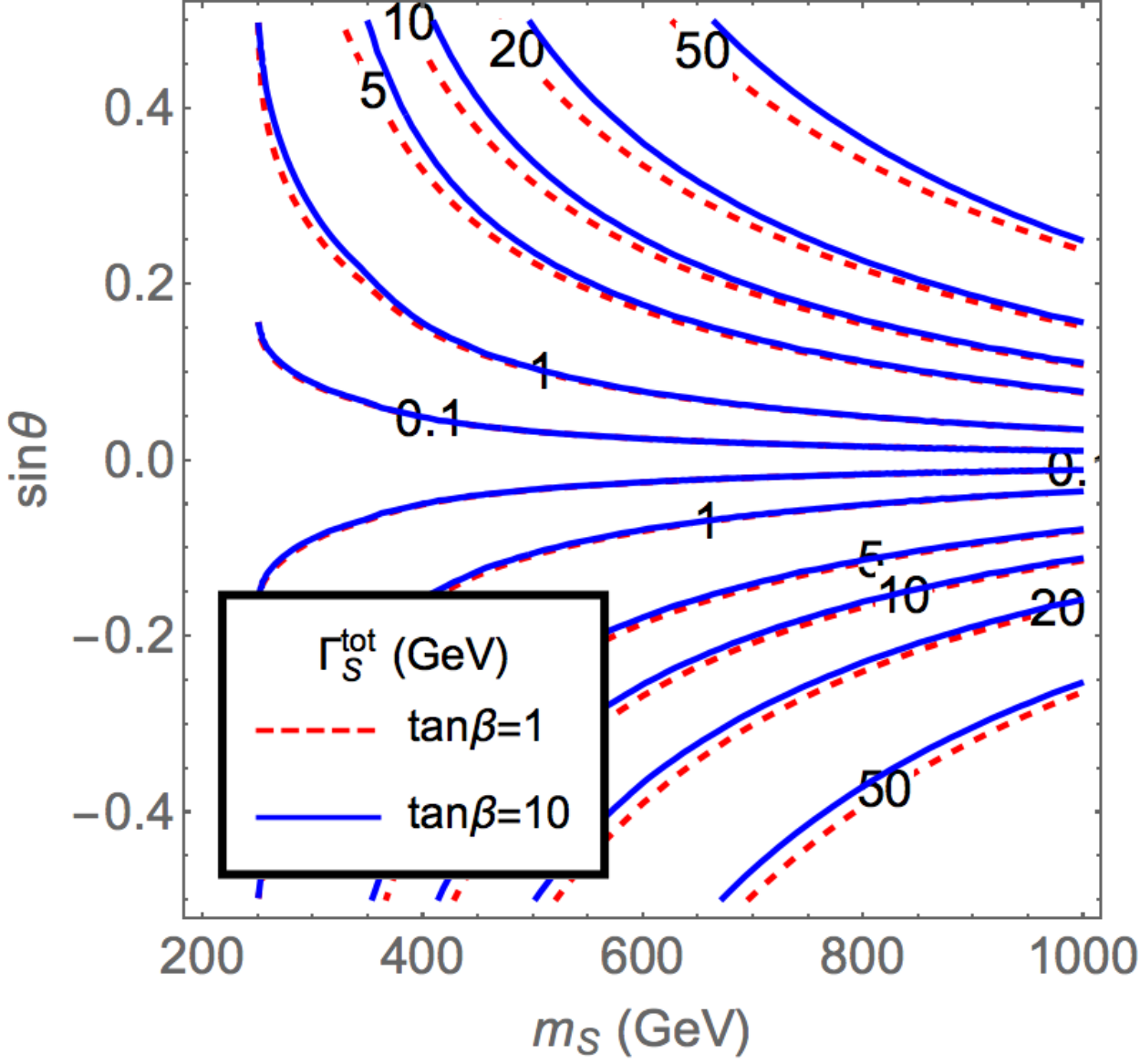} 
  \includegraphics[width=.48\textwidth]{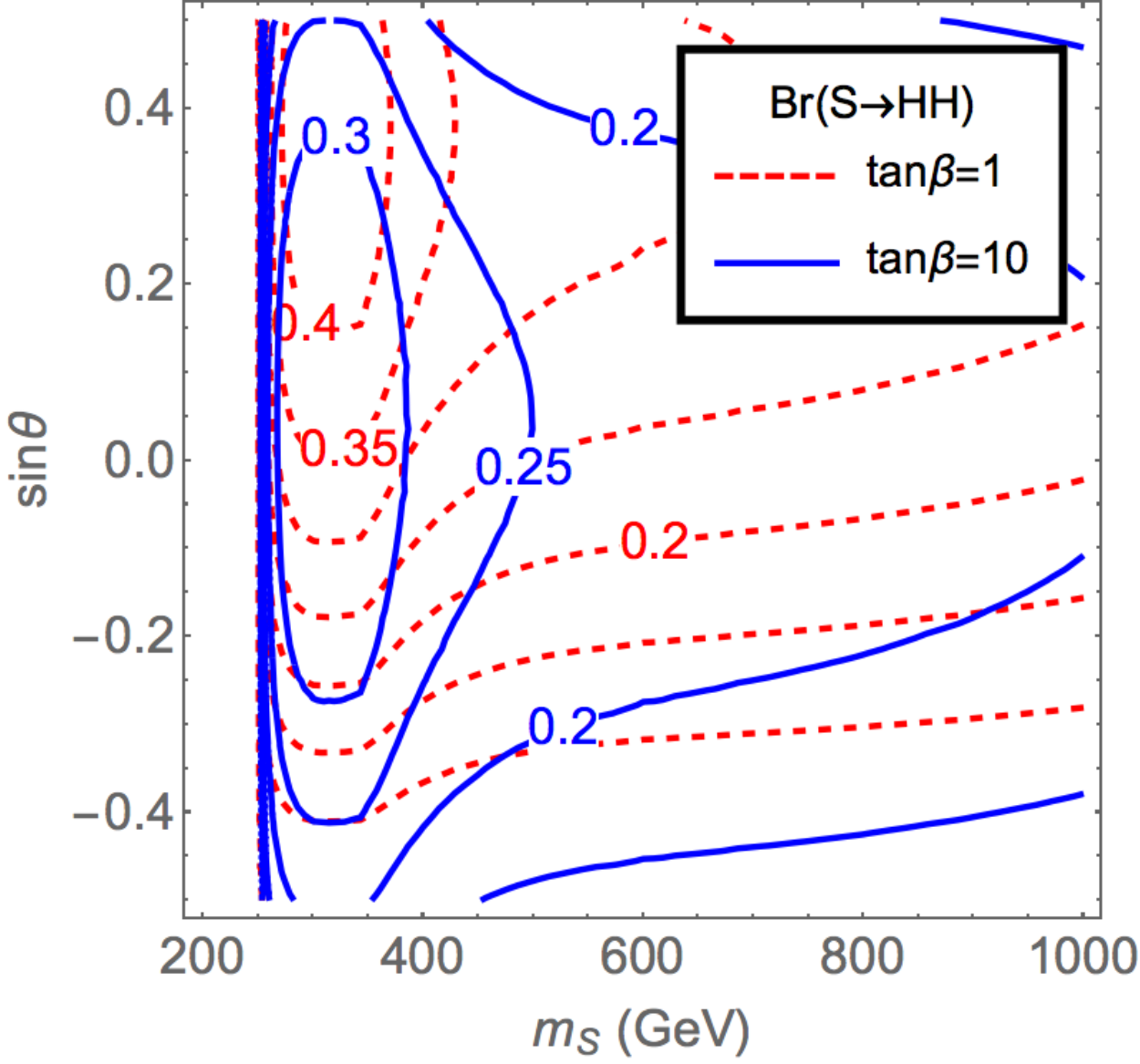}
  \caption{The total width (left panel)  and  branching fraction to Higgs pairs $\BR (S \rightarrow HH)$ (right panel)  of the singlet-like scalar $S$ as a function of the singlet-like scalar mass $m_S$ and the mixing angle $\sin\theta$. The red (dashed) contours and blue (solid) contours correspond to $\tan\beta$ = 1 and 10, respectively.
  %\ZL{We need to update the SM width part with threshold effect; for now we are using the asymptotic formulae for the width.}
  }
  \label{fig:scalardecay}
\end{figure} 

In the left panel of Fig.~\ref.{fig:scalardecay}, we show the total width of the heavy scalar state as a function of its mass and the mixing angle for $\tan\beta$ values of 1 and 10. We can see that its total width is not particularly sensitive to $\tan\beta$. %, while it grows to roughly to the third power of mass, as explained before. 
On the right panel of Fig.~\ref{fig:scalardecay}, we show the singlet decay branching ratio to Higgs pairs in the plane of the singlet scalar mass and the singlet-doublet mixing angle for $\tan\beta$ of 1 (red, dashed lines) and 10 (blue, solid lines), respectively. The branching fraction features a rapid decrease of roughly 5\% near the $t\bar t$ threshold due to the opening of this new decay channel. In addition, due to the possible cancellation from contributions to the $\lambda_{SHH}$ trilinear coupling in parameter space, as depicted in Eq.~(\ref{eq:211}), 
%different values of $\tan\beta$ show different contour shapes for constant branching fractions.
one can see strong variations in contour shapes for each value of $\tan\beta$.

The partial width of the singlet to Higgs bosons $\Gamma_S(S\to HH)$ scales as the third power of the scalar mass for a heavy scalar. This can be easily understood from Eq.~(\ref{eq:211}) and Eq.~(\ref{eq:gammaS}). %, where we can see that the width scales quadratically in $\lambda_{SHH}$ and the coupling grows also quadratically in $m_S$ for large $m_S$, giving a scaling $\sim m_S^3$. 
The partial width to $WW$ and $ZZ$ through the mixing with the SM Higgs also grows as the third power of the scalar mass due to the longitudinal enhancement for the massive vector gauge bosons. Consequently, the singlet branching fraction to $HH$ remains in the 20\% to 40\% range over a large span of the parameter space.

\section{ENHANCING THE DI-HIGGS SIGNAL VIA INTERFERENCE EFFECTS}
\label{sec:interference}

The on-shell interference effect may enhance or suppress the conventional Breit-Wigner resonance production. 
Examples in Higgs physics known in the literature, such as $gg\to h\to\gamma\gamma$~\cite{Campbell:2017rke} and $gg\to H\to t\bar t$~\cite{Carena:2016npr}, are both destructive.
We discuss in detail in this section the on-shell interference effect between the resonant singlet amplitude and the SM di-Higgs box diagram. We shall show that in the singlet extension of the SM considered in this paper, the on-shell interference effect is generically constructive and could be large in magnitude, thus enhances the signal production rate.

\subsection{Anatomy of the interference effect}

The interference effect between two generic amplitudes can be denoted as nonresonant amplitude $A_{nr}$ and resonant amplitude $A_{res}$.
The resonant amplitude $A_{res}$, defined as
\beq
A_{res} = a_{res} \frac {\hat s} {\hat s - m^2 + i \Gamma m},
\eeq
has a pole in the region of interest and 
we parametrize it as the product of a fast varying piece containing its propagator and a slowly varying piece $a_{res}$ that generically is a product of couplings and loop-functions. The general interference effect can then be parameterized as~\cite{Carena:2016npr,Campbell:2017rke},
\bea
|\mathcal{M}|_{int}^2 &=& 2\Re(A_{res}\times A_{nr}^*)\,=\,2\left(\mathcal{I}_{int} + \mathcal{R}_{int}\right),\nonumber \\
\mathcal{R}_{int} &\equiv& |A_{nr}||a_{res}|\frac {\hat s (\hat s - m^2)} {(\hat s - m^2)^2+\Gamma^2 m^2} \cos(\delta_{res}-\delta_{nr})\nonumber \\
\mathcal{I}_{int} &\equiv& |A_{nr}||a_{res}|\frac {\hat s \Gamma m} {(\hat s - m^2)^2+\Gamma^2 m^2} \sin(\delta_{res}-\delta_{nr}),
\label{eq:decomposition}
\eea
where $\delta_{res}$ and $\delta_{nr}$ denote the complex phases of $a_{res}$ and $A_{nr}$, respectively.

Schematically, the three amplitudes that enter the di-Higgs production can be parametrized as the following,
\bea
A_\vartriangleright^H(\hat s) &\propto& f_\vartriangleright(\hat s)\cos\theta~\frac {\lambda_{HHH}} {v} \frac {\hat s} {\hat s - m_H^2}\label{eq:samptriangle}\\
A_\square^H(\hat s) &\propto& f_\square(\hat s)\cos^2\theta\\
A_\vartriangleright^S(\hat s) &\propto& f_\vartriangleright(\hat s)\sin\theta~\frac{\lambda_{SHH}} {v} \frac {\hat s} {\hat s -m_S^2 + i \Gamma_S m_S},
\eea
where $f_\vartriangleright (\hat s)$ and $f_\square(\hat s)$ are the corresponding loop functions.
In Eq.~(\ref{eq:samptriangle}) we have dropped the nonimportant factors for the SM Higgs total width as the pair production is far above the SM Higgs on-shell condition.
For a CP-conserving theory that we are considering, all of the above parameters are real, except for the loop-functions $f_\vartriangleright(\hat s)$ and $f_\square(\hat s)$. The relevant phase between these loop functions\footnote{These loop functions can also be understood as form factors of the effective gluon-gluon-Higgs(-Higgs) couplings.} induces nontrivial interference effect between these diagrams.

\begin{figure}[t]  
  \centering
  \includegraphics[width=.48\textwidth]{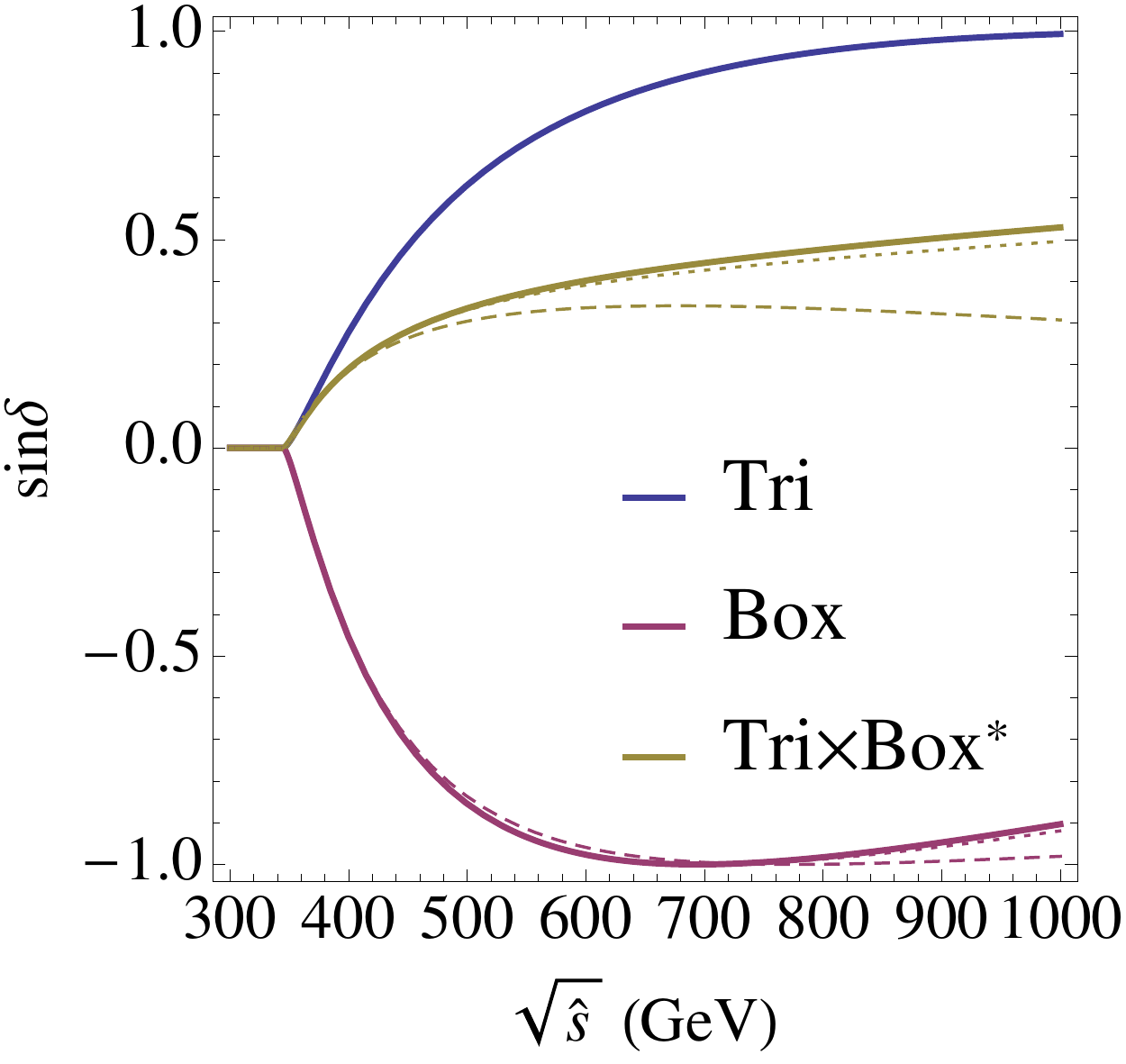}
  \caption{The phase of the interfering triangle and box amplitudes as a function of the partonic center of mass energy $\sqrt{\hat s}$. The solid, dotted, and dashed curves correspond to scattering angles of 0, 0.5 and 1 respectively. 
  %\ZL{planned only to use the left panel. The right panel is for our reference of the understanding of the asympotic behaivor of the phases. We can see the interesting splitting between the centeral and forward scattering amplitudes for the box background at large scale.}
  }
  \label{fig:phase} 
\end{figure}

The detailed expressions for these three amplitudes can be found in Ref.~\cite{Glover:1987nx}. The SM box contribution contains two pieces $f_\square$ and $g_\square$. The $g_\square$ piece corresponds to different helicity combinations of the gluons that does not interfere with the resonant term. 
In Fig.~\ref{fig:phase}, we show as a function of the partonic center of mass energy $\sqrt{\hat s}$, the phases of the triangle and box loop functions and their relative phase in blue, magenta and yellow curves, respectively. 
We observe that the phases of both diagrams start to increase after the $t\bar t$ threshold, as expected from the optical theorem. In particular, the relative phase between the interfering box and triangle diagrams, shown as the yellow curves, grows quickly after the threshold and remains large for the entire region under consideration. This relative strong phase drives the physics discussed in this paper as it allows for a nonvanishing $\mathcal{I}_{int}$ interference effect between the singlet resonance diagram and the SM box diagram.

\begin{table}[tbp]
  \label{tab:inter}
  \caption{Decomposition of all the allowed interference terms and their characteristics in the CP-conserving theory under consideration. The fourth column picks up the model-parameter dependence. The last column represents the sign of the interference term below/above the heavy scalar mass pole. The proportionality for $\mathcal{I}_{int}$ of the SM piece denoted $0^*$ contains more factors than the model parameter; see details in the text. 
  %The $\mathcal{I}_{int}$ interference piece are the focus of this study, which do not show up in various places and thus often overlooked in my studies. 
  }
  \begin{center}
  \begin{tabular}{|c|c|c|c|c|}
  \hline
 \multicolumn{2}{|c|}{Inter. Term.}
 & Rel. phase
 & Proportionality
 & Inter. sign\\ \hline \hline
 \multirow{2}{*}{$A^H_\vartriangleright$-$A^H_\square$}
  & $ \mathcal{R}_{int} $
  & $\cos(\delta_\vartriangleright-\delta_\square)$
  & $\cos^3\theta \lambda_{HHH}$
  & $-$
  \\ \cline{2-5}
  & $\mathcal{I}_{int}$
  & $\sin(\delta_\vartriangleright-\delta_\square)$
  & $0^*$
  & \color{red} 0
   \\ \hline \hline
\multirow{2}{*}{$A^S_\vartriangleright$-$A^H_\vartriangleright$}
  & $ \mathcal{R}_{int} $
  & 1 
  & $\lambda_{SHH} \lambda_{HHH} \cos\theta~\sin\theta$
  & $-/+$
  \\ \cline{2-5}
  & $\mathcal{I}_{int}$
  & 0
  & $\lambda_{SHH} \lambda_{HHH} \cos\theta~\sin\theta$
  & \color{red} 0
  \\ \hline \hline
\multirow{2}{*}{$A^S_\vartriangleright$-$A^H_\square$}
  & $ \mathcal{R}_{int} $
  & $\cos(\delta_\vartriangleright-\delta_\square)$ 
  & $\lambda_{SHH} \cos^2\theta \sin\theta$
  & $+/-$
  \\ \cline{2-5}
  & $\mathcal{I}_{int}$ 
  & $\sin(\delta_\vartriangleright-\delta_\square)$
  & $\lambda_{SHH} \cos^2\theta \sin\theta$
  & $\color{red} +$
  \\ 
  \hline
  \end{tabular}
  \end{center}
  \label{tab:int}
  \vspace*{-0.6cm}
\end{table}%

In Table~\ref{tab:int}, we summarize the different behaviors of all the interference terms allowed in this theory. We decompose the interference effects into the $\mathcal{R}_{int}$ and $\mathcal{I}_{int}$, as defined in Eq.~(\ref{eq:decomposition}), and further highlight their dependence on the relative phase, model parameters and the resulting signs of the interference effects. 

The special $\mathcal{I}_{int}$ terms vanish both for the interference between the SM diagrams $A^H_\vartriangleright$-$A^H_\square$ and the interference between the resonant singlet and the SM triangle diagrams $A^S_\vartriangleright$-$A^H_\vartriangleright$ for different reasons. 
For the latter, the singlet resonant amplitude and the SM triangle amplitude share a common source of the strong phase $\delta_\vartriangleright$ from the triangle fermionic loop of the induced gluon-gluon-scalar coupling. Hence, $\sin(\delta_\vartriangleright -\delta_\vartriangleright)=0$ and this makes Eq.~(\ref{eq:decomposition}) vanish.
%, which put these two amplitudes exactly in phase apart from the resonant propagator. 
For the interference between the SM box and triangle diagrams $A^H_\vartriangleright$-$A^H_\square$, denoted $0^*$ in the proportionality column of the table, the $\mathcal{I}_{int}$ vanishes because we are always in the off-shell regime for the intermediate SM Higgs in the triangle diagram.
%Both the triangle and box diagram apart from the loop-functions are purely real, and consequently there are no contributions proportional to the (relative) imaginary part from their loop functions. One can also view this from a slightly different perspective,
Viewing the SM triangle diagram as $A_{res}$, then the $\mathcal{I}_{int}$ part in Eq.~(\ref{eq:decomposition}) is strictly nonzero. However, due to the fact that we can never hit the SM Higgs pole in the relevant regime $\hat s > (2 m_H)^2$, such contribution is 
\beq
\frac {\hat s \Gamma_H m_H} {(\hat s - m_H^2)^2+\Gamma_H^2 m_H^2} < \frac {4 m_H^3 \Gamma_H} {9 m_H^4}\approx 1.5\times 10^{-5},
\eeq
and hence can be neglected.

In contrast, the special interference effect $\mathcal{I}_{int}$ only appears between the singlet resonant diagram and the SM box digram $A^S_\vartriangleright$-$A^H_\square$. This interference effect is proportional to the relative phase between the loop functions $\sin(\delta_\vartriangleright-\delta_\square)$ and the imaginary part of the scalar propagator which is sizable near the scalar mass pole. In this work, we pay special attention to this effect whose importance has been overlooked in the past literature. 

The signs of the interference effects are determined by a product of relative phases, model parameters and kinematics. The relative phases are always positive for the mass range considered here, as shown in Fig.~\ref{fig:phase}. The kinematics straightforwardly relies on the Higgs pair invariant mass with respect to the heavy scalar mass pole. %In addition, for the spontaneously $Z_2$ breaking scenario considered here, the signs of the interference effects end up being fixed as shown in \autoref{tab:int}. 
%The coupling product from the model parameters are fixed due to model construction and consistency requirement that, $-\lambda_{SHH}$ and $\sin\theta$ share a common sign, following the sign of $\lambda_{s\phi}$ in the original potential.
In addition due to spontaneously $Z_2$ breaking model construction and consistency requirement, $\lambda_{SHH}$ and $\sin\theta$ has opposite signs, in accordance to the sign of $\lambda_{s\phi}$ in the original potential.
%For the spontaneously $Z_2$ breaking scenario considered here, 
The overall signs of the interference effects end up being fixed as shown in Table~\ref{tab:int}. 

\subsection{Parametric dependence of the on-shell interference effect}

\begin{figure}[t]  
  \centering
  \includegraphics[width=.48\textwidth]{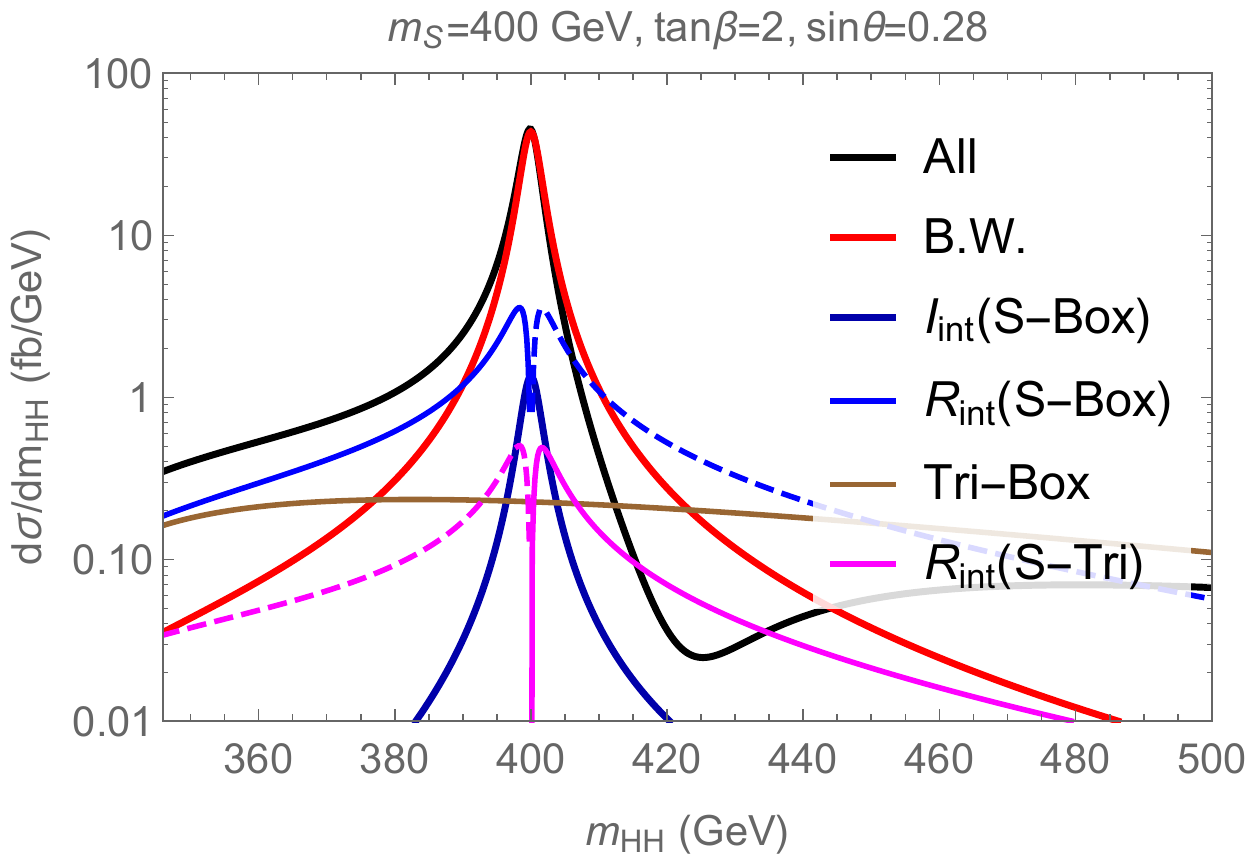}
  \includegraphics[width=.49\textwidth]{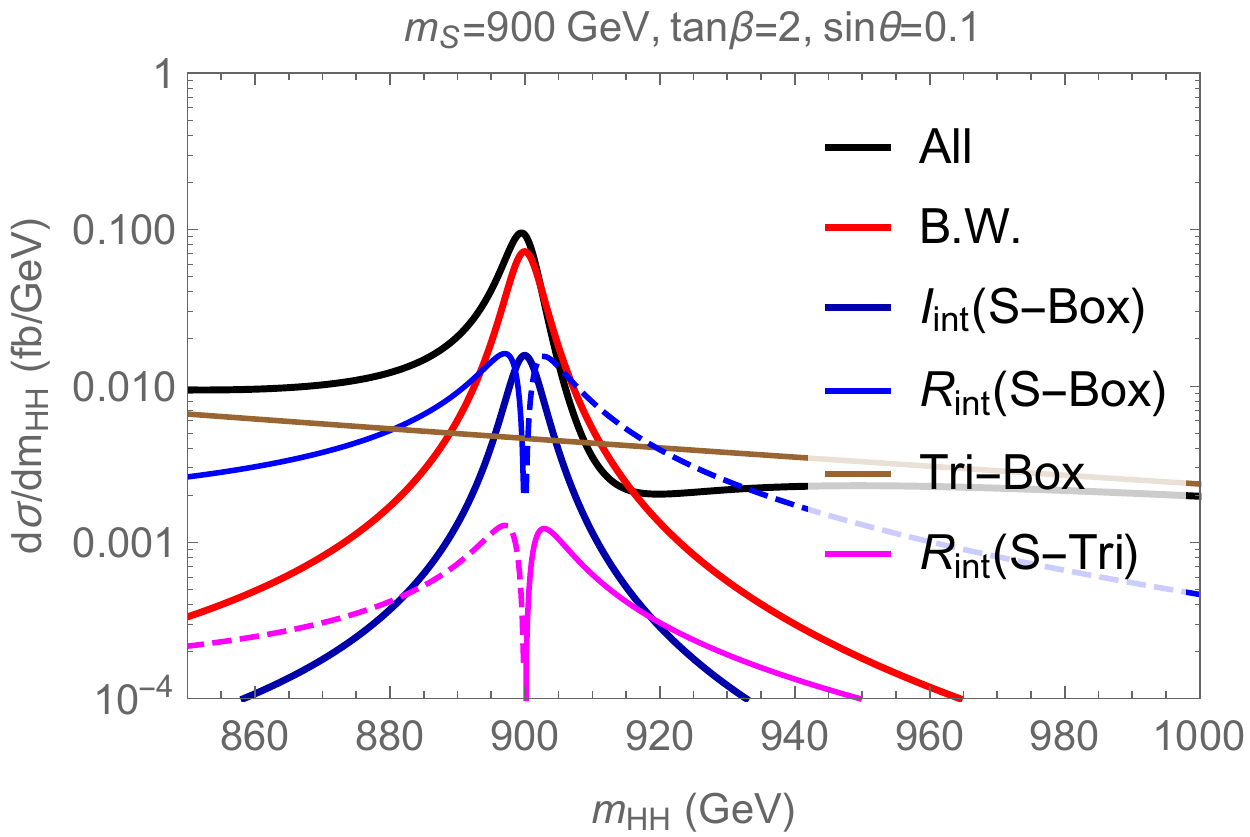}
  \caption{Decomposition of the differential distribution of the Higgs pair production in presence of a singlet resonance at 13 TeV LHC. The black curve represents the overall line shape after coherent sum of all amplitudes squared. The red curve represent the Breit-Wigner resonance piece from the singlet resonant production. The dark blue (thick) curve represents the novel interference term between the singlet resonant amplitude and the SM box amplitude that enhances the signal resonant production, noting the identical line shape of this contribution to that of the Breit-Wigner piece in red curves. The blue, brown and magenta lines represent the conventional interference terms $\mathcal{R}_{int}$ between the three amplitudes. We show the corresponding destructive interference effects in dashed curves. 
  %\ZL{Change the left panel to tanbeta=2 to match Daswon and Lewis.}
  %\ZL{y-axis normalization need to be updated/corrected.}
  }
  \label{fig:decomposition} 
\end{figure}

After understanding the sources of various interference effects, especially the on-shell interference effect $I_{int}$, we study its parametric dependence in this section.\footnote{Throughout this work we use the finite $m_t$ result at leading order~\cite{Glover:1987nx}. We adopt the K-factor between the next-to-leading-order and the leading-order result in the $m_{hh}$ distribution provided by Ref.~\cite{Dawson:2015haa}.}

We first show the line-shape decomposition into components discussed in Table~\ref{tab:int} for two benchmark points in Fig.~\ref{fig:decomposition}. We display the Breit-Wigner, nonresonant line shape from SM triangle and box diagrams, and the total line shape in red, brown and black curves, respectively. The interference terms $\mathcal{R}_{int}$ proportional to the real part of the heavy scalar propagator are shown in blue and magenta curves. Observe that these interference terms flip their signs when crossing the scalar mass pole and this is shown by switching the solid curve for constructive interference to dashed curves for destructive interference.
%the destructive interference are shown in dashed curves. 
Finally, for the interference term proportional to the imaginary component of the scalar propagator, we show the special term $I_{int}$ in (thick) dark blue curve. We can observe that the $I_{int}$ piece has very similar line shape to the Breit-Wigner resonance piece near the scalar mass pole. We shall denote this term $\mathcal{I}_{int}$ as on-shell interference effect, since $\mathcal{I}_{int}$ acquires its maximal value precisely on-shell. This is in contrast to the term $\mathcal{R}_{int}$ that vanishes when the invariant mass of the final state is precisely at the scalar mass pole.

In the left panel of Fig.~\ref{fig:decomposition}, we choose as a benchmark $m_S=400$ GeV, $\tan\beta=2$ and mixing angle $\sin\theta=0.28$ to match one of the benchmarks in Ref.~\cite{Dawson:2015haa}.\footnote{Our definition of $\tan\beta$ is the inverse of the $\tan\beta$ definition used in Ref.~\cite{Dawson:2015haa}.} For this benchmark, we reproduce their result on the overall line shape and some of the specific line shape contributions shown in Ref.~\cite{Dawson:2015haa}.
%Notably, we emphasize on the on-shell interference term $I_{int}$ shown in (thick) dark blue curve. 
For this benchmark, the on-shell interference term $\mathcal{I}_{int}$, shown in the (thick) dark blue curve, is smaller than the Breit-Wigner contribution by almost two orders of magnitude and thus can be neglected. Instead, in the right panel of Fig.~\ref{fig:decomposition}, we show the line shape decomposition for a different benchmark point of heavy scalar mass $m_S=900$ GeV, $\tan\beta=2$ and mixing angle $\sin\theta=0.1$. 
%This choice is to have similar width as in the left-panel. 
For this benchmark, we can clearly observe the contribution from the on-shell interference term $\mathcal{I}_{int}$, as its magnitude is more than 15\% of the Breit-Wigner resonance shown by the  red curve. This leads to an enhancement when comparing the overall line shape (black curve) to the Breit-Wigner resonance alone near the resonance peak.

\begin{figure}[t]
  \centering
  \includegraphics[width=1\textwidth]{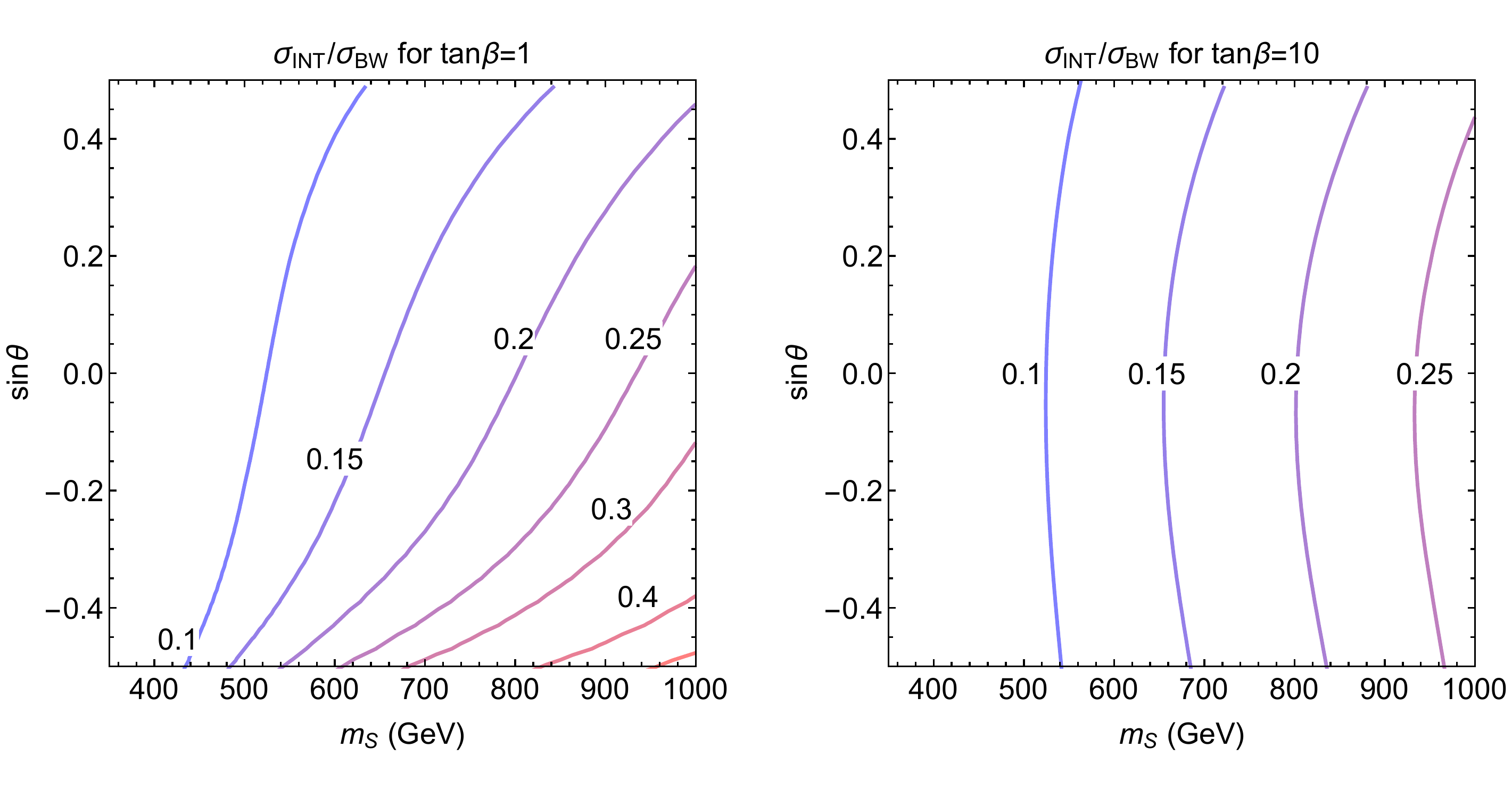} 
  \caption{The relative size of the on-shell interference effect with respect to the Breit-Wigner contribution for the scalar singlet resonant production after averaging over the scattering angle $\cos\theta^*$ from $-0.5$ to +0.5 for central scattering.
  %The left, middle and right panel corresponds to $\tan\beta$ values of 1, 2 and 10 respectively. The shaded region are disallowed by vacuum stability and perturbative unitarity argument.
  }
  \label{fig:interference_central}
\end{figure} 

With the comprehensive understanding of the interference effect, we can quantify the relative size of the on-shell interference effect by normalizing it to the the Breit-Wigner contribution,
%, namely the cross section ratios of the interference term near the pole over the Breit-Wigner term 
$\sigma_{\rm Int}/\sigma_{\rm B.W.}$. 
This ratio is well defined due to the similar line shapes of these two contributions near the mass pole. 
%Due to the angular dependence of the interfering box amplitude, 
We integrate over the scattering angle in the center of mass frame in the $-0.5$ to $+0.5$ range for central scattering and average over the ratio. We show in Fig.~\ref{fig:interference_central} the parametric dependence of this interference effect as a function of the heavy scalar mass $m_S$ and singlet-doublet mixing angle $\sin\theta$ for $\tan\beta$ = 1  (left panel) and $\tan\beta$ = 10 (right panel). 
%From \autoref{fig:interference_central} 
We obtain that  the size of this on-shell interference effect $I_{int}$ varies between a few percent to up to 40\% of the size of the Breit-Wigner resonance for the parameter region considered in this study. The effect is further enhanced for heavier scalar masses and larger widths. The quantitative differences of the iso-curvatures between the two panels  in Fig.~\ref{fig:interference_central} are caused by the parametric dependence of $\lambda_{SHH}$  and the singlet total decay width shown in Fig.~\ref{fig:trilinear} and Fig.~\ref{fig:scalardecay}, respectively. Clearly, the interference effect could play an important role in the phenomenology and further determination of model parameters if the heavy scalar is discovered.

%\subsection{Parametric dependence of the on-shell interference effect for $ZZ$ and $WW$ channel}

\section{PHENOMENOLOGICAL STUDY}%Phenomenological study}
\label{sec:pheno}

We present in this section our analysis of the differential distribution of the Higgs pair invariant mass to estimate the relevance of the interference effects discussed in the previous section. We choose one of the best channels, $pp\to HH \to b\bar b \gamma\gamma$, as the benchmark channel to present the details of our analysis. Furthermore, we discuss another phenomenologically relevant piece of interference in the far off-shell region of the singlet scalar. We display the discovery and exclusion reach  for both HL-LHC and HE-LHC  for various values of $\tan\beta$ in the $m_S$-$\sin\theta$ plane. Finally in the last part of this section, we discuss the relevance of the di-Higgs channel in probing the strength of the first-order electroweak phase transition in a simplified effective field theory (EFT) approach for both the spontaneous $Z_2$ breaking scenario and an explicit $Z_2$ breaking scenario.

\subsection{Differential distribution}

\begin{figure}[t]  
  \centering
  \includegraphics[width=.60\textwidth]{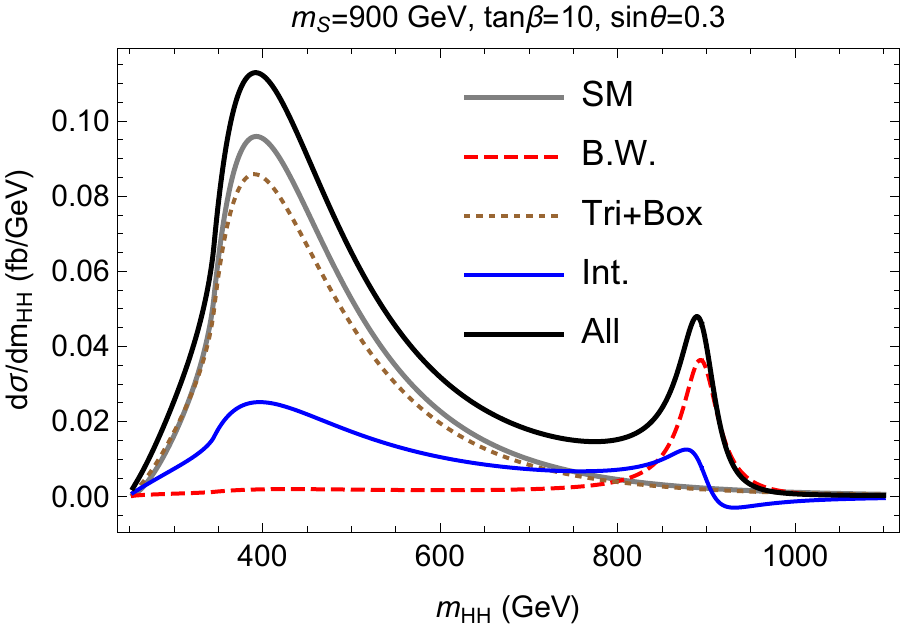}
  \caption{
  The differential di-Higgs distribution for a benchmark point of the singlet extension of the SM shown in linear scale and over a broad range of the di-Higgs invariant mass. The full results for the SM and the singlet SM extension  are shown by the  gray and black curves, respectively. In the singlet extension of the SM, the contributions from the resonant singlet diagram, the nonresonant diagram and the interference between them are shown in red (dashed), brown (dotted) and blue curves, respectively.  
  %Decomposition of the differential distribution of the Higgs pair production in presences of a singlet resonance. The black curve represent the overall lineshape after coherent sum of all amplitudes squared. The red curve represent the Breit-Wigner resonance piece from the singlet resonant production. The dark blue (thick) curve represent the novel interference term between the singlet resonant amplitude and the SM box amplitude that enhances the signal resonant production, noting the identical lineshape of this contribution to that of the Breit-Wigner piece in red curves. The blue, brown and magenta lines represent the conventional interference terms $\mathcal{R}_{int}$ between the three amplitudes. We show the corresponding destructive interference effects in dashed curves. 
  %\ZL{y-axis normalization need to be updated/corrected.}
  }
  \label{fig:phenoshape} 
\end{figure}

%%Before considering a full detailed analysis of interference effects and the projected results, we would like to comment on the phenomenologically important off-shell interference effect.
%, in addition to the on-shell interference effect around the pole of the heavy scalar. 
In Fig.~\ref{fig:phenoshape} we display the differential cross section as a function of the Higgs pair invariant mass for a benchmark point with a heavy scalar mass of 900 GeV, mixing angle $\sin\theta=0.3$ and $\tan\beta=10$. 
%Instead of showing the cross section in logarithmic scale and in the vicinity of the heavy scalar mass pole as in \autoref{fig:decomposition}, we show from a more observational perspective for the cross section in linear scale and a large range of the Higgs pair invariant mass, covering the low invariant mass regime favor by parton distribution functions at hadron colliders.
The differential cross section is shown in linear scale for a broad range of di-Higgs invariant masses,  including the low invariant mass regime favored by parton distribution functions at hadron colliders.

We choose this benchmark to show well the separation of the scalar resonance peak and the threshold enhancement peak above the $t\bar t$-threshold. The SM Higgs pair invariant mass distribution is given by the  gray curve while the black curve depicts the di-Higgs invariant mass distribution from the singlet extension of the SM. 
It is informative to present all three pieces that contribute to the full result of the di-Higgs production, namely, the resonance contribution (red, dashed curve), the SM nonresonance contribution (box and triangle diagrams given by the brown, dotted curve), and the interference between them (blue curve). 
Note that the small difference between the ``Tri+Box'' and the ``SM'' line shapes is caused by the doublet-singlet scalar mixing, which leads to a $\cos\theta$ suppression of the SM-like Higgs coupling to top quarks as well as a modified SM-like Higgs trilinear coupling $\lambda_{HHH}$, as depicted Eq.~(\ref{eq:111}).
We observe that the full results show an important enhancement in the di-Higgs production across a  large range of invariant masses. This behavior is anticipated from the decomposition analysis in the previous section. There is a clear net effect from  the interference curve shown in blue.  Close to the the scalar mass pole at 900 GeV, the on-shell interference effect enhances the Breit-Wigner resonances peak (red, dashed curve) by about 25\%. Off-the resonance peak, and especially at the threshold peak, the interference term (blue curve) enhances the cross section quite sizably as well. Hence, a combined differential analysis in the Higgs pair invariant mass is crucial in probing the singlet extension of the SM. 
%Moreover, a consistent treatment is required not only for the purpose of exclusion but also in extraction of the physics information contained in potential deviations. 

\subsection{Signal and background analysis for $pp\to HH \to b\bar b \gamma\gamma$}

\begin{table}[tbp]
  \label{tab:inter}
  \caption{Summary of expected number of events for the SM Higgs pair production and the SM backgrounds for the $b\bar b \gamma\gamma$ di-Higgs search after selection cuts, obtained from Ref.~\cite{Azatov:2015oxa} for the HL-LHC and further extrapolated for the HE-LHC.
  }
  \begin{center}
  \begin{tabular}{|c|c|c|c|c|c|c|c|}
  \hline
 \# of events
 & \multicolumn{2}{|c|}{HL-LHC} 
 & \multicolumn{2}{|c|}{HE-LHC}
 \\
 expected & \multicolumn{2}{|c|}{13 TeV @ $3~\abi$} & \multicolumn{2}{|c|}{27 TeV @ $10~\abi$}
 \\ \hline \hline
 bins (GeV)
 & SM HH
 & SM BKG
 & SM HH
 & SM BKG
 \\ \hline
250--400   & 2.1  & 12.0 & 33.2  & 186.4 \\ \hline
400--550   & 6.3  & 15.9 & 110.9 & 278.8 \\ \hline 
550--700   & 2.9  & 5.2  & 58.4  & 105.6 \\ \hline
700--850   & 1.0  & 2.0  & 23.4  & 46.7  \\ \hline
850--1000  & 0.3  & 1.4  & 8.9   & 38.8  \\ \hline
1000--1200 & 0.2  & 0.7  & 4.7   & 20.4  \\ \hline
1200--1400 & --    & --    & 1.9   & 8.0   \\ \hline
1400--1600 & --    & --    & 0.8   & 3.5   \\ \hline
1600--1800 & --    & --    & 0.4   & 1.7   \\ \hline
1800--2000 & --    & --    & 0.2   & 0.9   \\ \hline
\hline
  \end{tabular}
  \end{center}
  \label{tab:binning}
  \vspace*{-0.6cm}
\end{table}%

In the following, we consider the di-Higgs decaying into $b\bar b \gamma\gamma$  in the singlet extension of the SM, and perform a consistent treatment of the interference effect and a differential analysis of the line shapes. 
Although this channel is one of the most sensitive ones due to its balance between the cleanness of the final state and the signal statistics,  the detailed analysis is nevertheless quite involved. For both the SM signal and background expected number of events at HL-LHC, we use the simulated and validated results listed in Table~V of Ref.~\cite{Azatov:2015oxa}. To extrapolate the signal expected from our singlet extension of the SM, we assume the same acceptance as the SM Higgs pair. For HE-LHC with a center of mass energy of 27 TeV, we assume the same acceptance as the HL-LHC that varies between 10\% to 30\% for the di-Higgs signal.
For the SM background at the HE-LHC, we assume the same signal to background ratio as the HL-LHC in the low invariant mass bins, while for the high invariant mass bins we consider a fixed signal to background ratio of 23\%.
%{\textcolor{blue} { For the SM background in the high invariant mass bins, we further assume a fixed di-Higgs signal over background ratio of $23\%$   ???}}\ZL{, which is the average signal background ratio for $m_{HH}>850$ GeV in the HL-LHC simulation}. 
In Table~\ref{tab:binning}, we tabulate the expected number of events for the SM Higgs pair and SM background.\footnote{Although this analysis includes different signal efficiencies depending on different Higgs pair invariant mass windows, a future analysis focusing in high invariant mass bins could lead to improved results, especially when combined with different decay final states.}
% using the extrapolation method detailed above.

We calculate and combine the significance of each bin using the following approximation~\cite{Cowan:2010js},
\beq
\Delta\chi^2 = \sum_i^{\rm bins} 2\left((n_{s,i}+n_{b,i})\log(1+\frac {n_{s,i}} {n_{b,i}})-n_{s,i}\right),
\eeq
assuming all the bins are independent. As shown in Table~\ref{tab:binning}, the bins are typically with low statistics, therefore it is reasonable to ignore systematics at this stage. We assume that the observed number of events in this channel follows the SM expectation values. $n_{b,i}$ represents the sum of the SM di-Higgs event rate and its background for each mass window listed in Table~\ref{tab:binning}; $n_{s,i}$  represents the difference generated from the singlet model in the di-Higgs production channel with respect to the SM Higgs pair production in each bin. As shown in Ref.~\cite{Cowan:2010js}, this formulae provides a good approximation for the median discovery significance for a large range of underlying statistics, including relatively low statistical bins where Gaussian approximation fails.\footnote{Although still facing sizable differences for the true significance with statistical simulations~\cite{Cowan:2010js}, the above treatment is sufficient for our current study as our purpose is to demonstrate the impact of the interference effect.}

%\ZL{add K-factor discussion and references in one or two sentences or earlier when we first show the lineshapes.}
%\ZL{Need to comment on the exclusion and discovery potential from other channels, such as $ZZ$ and $WW$ resonance searches in relevant places.}

\subsection{Discovery and exclusion reach of the HL- and HE-LHC} 

\begin{figure}[t]
  \centering
  \includegraphics[width=1\textwidth]{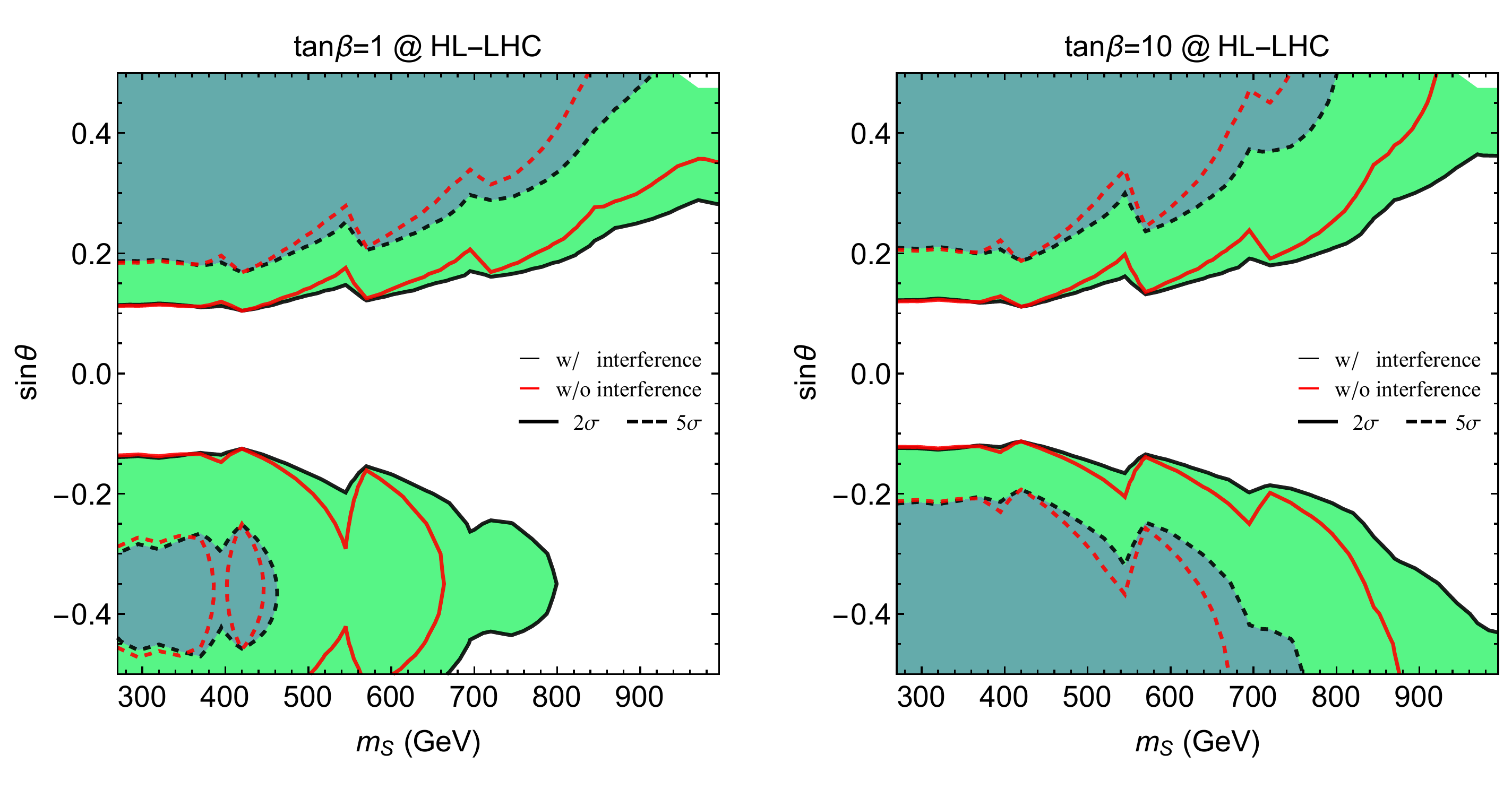}
  \caption{Projected exclusion and discovery limits at HL-LHC in the $m_S$-$\sin\theta$ plane with the line-shape analysis detailed in the text for $\tan\beta=1$ (left panel) and $\tan\beta=10$ (right panel). The shaded regions bounded by dashed/solid curves are within the discovery/exclusion reach of the HL-LHC. The black and red lines represent the projection with and without the inclusion of the interference effects between the singlet resonance diagram and the SM Higgs pair diagram, respectively.
  %The left, middle and right panel corresponds to $\tan\beta$ values of 1, 2 and 10 respectively. The shaded region are disallowed by vacuum stability and perturbative unitarity argument.
  }
  \label{fig:HLprojection}
\end{figure} 

Using the analysis detailed above, we obtain the discovery and exclusion projections for the HL-LHC and HE-LHC.
In Fig.~\ref{fig:HLprojection} we show the projected 2-$\sigma$ exclusion and 5-$\sigma$ discovery reach for the HL-LHC in the $m_S$-$\sin\theta$ plane for $\tan\beta=1$ (left panel) and $\tan\beta=10$ (right panel) in solid and dashed curves, respectively. The shaded regions are within the reach of the HL-LHC for discovery and exclusion projections. To demonstrate the relevance of the interference effects discussed in the previous sections, we show both the results obtained with and without the inclusion of the interference effects in black and red contours, respectively. 
%We further overlay the disallowed regions from perturbative unitarity requirement and the EWPO constraints in gray and brown shaded regions, respectively.

We observe in Fig.~\ref{fig:HLprojection} that the inclusion of the interference effects extend the projections in a relevant way. For example, considering the $\tan\beta=10$ case in the right panel for $\sin\theta\simeq 0.35$ the interference effect increase the exclusion limit on $m_S$ from 850~GeV to 1000~GeV.
%, due to the large on-shell interference effects shown in the right panel of \autoref{fig:interference_central}. 
%Note that the perturbative unitarity does not constrain the parameter space of heavy scalar mass $m_S$ below a few TeV, the inclusion of the interference effect increases the discover and exclusion limit further in the sizable $\sin\theta$ region.
Note that the on-shell interference effect is larger for heavier scalar mass $m_S$. 
%, which in the cases of large $\tan\beta$ are still compatible with the perturbative unitarity requirement. 

\begin{figure}[t]
  \centering
\includegraphics[width=1\textwidth]{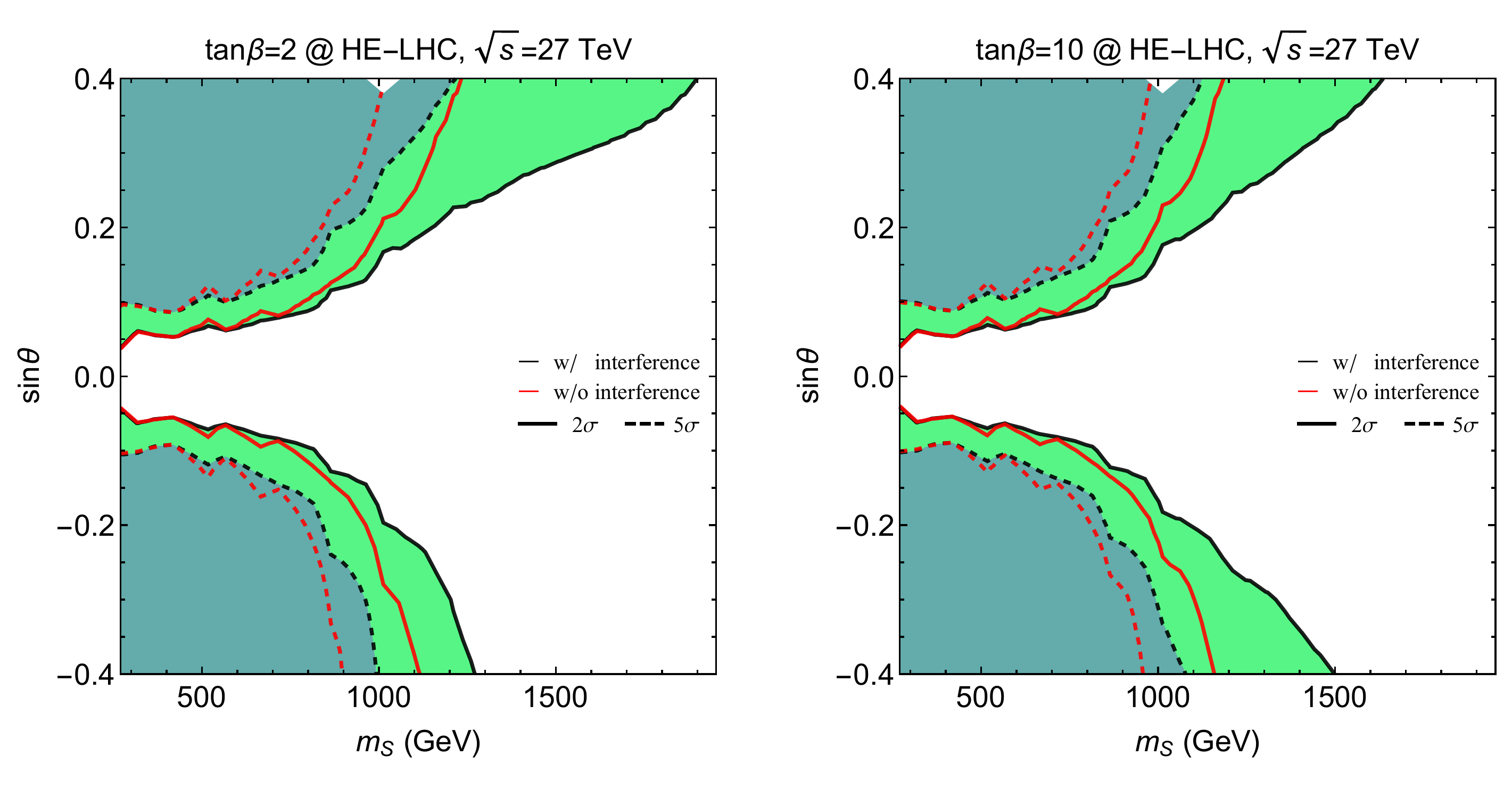}
  \caption{
    %\ZL{figure to be updated upon new data. }
  Similar to Fig.~\ref{fig:HLprojection}, projected exclusion and discovery limits at HE-LHC with 27 TeV center of mass energy and an integrated luminosity of 10~$\abi$ for $\tan\beta=2$ (left panel) and $\tan\beta=10$ (right panel).
  %The left, middle and right panel corresponds to $\tan\beta$ values of 1, 2 and 10 respectively. The shaded region are disallowed by vacuum stability and perturbative unitarity argument.
  }
  \label{fig:HEprojection}
\end{figure} 

In Fig.~\ref{fig:HEprojection} we show the projections for the HE-LHC in a analogous fashion as in Fig.~\ref{fig:HLprojection}. The discovery and exclusion reach for heavy scalars can be significantly extended by the HE-LHC operating at 27 TeV center of mass energy with 10 $\abi$ of integrated luminosity. We show the results for $\tan\beta=2$ (left panel) and $\tan\beta=10$ (right panel). 
For example, considering the $\tan\beta=2$ case in the right panel of Fig.~\ref{fig:HEprojection}, for $\sin\theta\simeq 0.35$ the exclusion reach increases from 1200 to 1800~GeV, once more showing the importance of including the on-shell interference effects.

In Sec.~\ref{sec:stabilityunitarityandEWSBconditions}, we have shown that in the spontaneous $Z_2$ breaking model, the perturbative unitarity requirement can place stringent upper bounds on the singlet scalar mass, depending on the value of $\tan\beta$. Such bounds are 
% rules out singlet scalar mass beyond 900 GeV in the spontaneous $Z_2$ breaking model under consideration. 
%This is 
driven by the large singlet quartic $\lambda_S$ needed to obtain heavy mass values from a relatively small vev $v_s=v\tan\beta$. 
In an explicit $Z_2$ breaking model, instead, larger values of the singlet mass are perfectly compatible with perturbative unitarity requirement even for small value of $\tan\beta$. 
%Therefore, we choose not to show the perturbative unitarity restrictions in \autoref{fig:HLprojection} and \autoref{fig:HEprojection}.
Therefore, in Fig.~\ref{fig:HLprojection} and Fig.~\ref{fig:HEprojection} we perform a general analysis for the LHC reach without imposing the perturbative unitarity restrictions.
%Still, instead of showing the limit for $\tan\beta$ of 1 and 10 as in \autoref{fig:HLprojection}, we show the results obtained for $\tan\beta$ of 2 and 10 in the left panel and right panel of \autoref{fig:HEprojection}, respectively.
%The derived limits shown in this figure can be viewed as in such an $Z_2$ explicit breaking model, which illustrates the relevance of the interference effect as the focus of this study. %We can see for $\sin\theta=0.4$, the inclusion of the interference effect significantly improves the exclusion limits from 1.2 TeV to 1.6 TeV for $\tan\beta=2$ and moderate improvement for both the discovery and exclusion reach in other cases. 
%However, for negative $\sin\theta$, the interference effect reduces the exclusion limits, this is
%\ZL{Discuss $\tan\beta=10$ resutls when numbers are ready.}

%Before discussing the relevance of the probed parameter region to first order electroweak phase transition in the next section. 
%There are several common aspects of the HL- and HE-LHC analysis worth mentioning. 
%First, a
%As discussed at the very beginning of this section, the purpose of this analysis is to demonstrate the impact of the interference contributions in the search for BSM effects in the di-Higgs final state. 
It is worth mentioning that when the heavy scalar resonance is divided evenly between two bins, its significance is reduced. This, together with a very coarse binning we choose in Table~\ref{tab:binning}, leads to the wiggles in the discovery and exclusion projection contours in this section. 
A more refined analysis that leads to smoother projections would be desirable.
%Although a more advanced treatment to obtain smooth projections would be dable
In addition, due to the mixing with the SM Higgs, the heavy scalar also has sizable branching fractions into $WW$ and $ZZ$, as implied in Fig.~\ref{fig:scalardecay}. New channels such as $gg\to S\to WW, ZZ$ could provide complementary and even competitive information and discovery potential for the heavy scalar. Note that similar on-shell and off-shell interference effects will take place in these channels as well.
%, see e.g., discussion in Ref.~\cite{deFlorian:2016spz}. 
Hence, it would be interesting to consider a comprehensive treatment and comparison between different search channels, such as other decays of the Higgs pair, as well as other decay modes of the heavy scalar. We reserve these for future study.

\subsection{Implications for the first-order electroweak phase transition}

In this section, we investigate the implications of the interference effects for the parameter regions enabling a first-order electroweak phase transition. 
%We discuss in this section. 
There are several phenomenological studies in the literature that investigate different realizations of first-order electroweak phase transitions in singlet extensions of the SM.
For the case of the $Z_2$ symmetric singlet extension, there are detailed studies in Ref.~\cite{Espinosa:2011ax,Curtin:2014jma,Curtin:2016urg}, including the possibilities of both 1-step and 2-step phase transitions. For a general singlet extend SM, several numerical and semi-analytical studies have been carried out~\cite{No:2013wsa,Huang:2015tdv,Chen:2017qcz}. 
%A detailed finite-temperature thermal history study is beyond the scope of this paper. 
Here we perform a simplified EFT analysis on the spontaneous $Z_2$ breaking scenario and a particular explicit $Z_2$ breaking scenario to illustrate the relevance of the interference effect.
A detailed finite-temperature thermal history study for the full theory will be presented elsewhere.
% and discuss the physics cases not well-approximated by this EFT analysis. 

A deformation of the Higgs thermal potential is the key to change the electroweak phase transition from second order to first order. The simplest way in EFT is to introduce the dimension-six operator $O_6\equiv (\phi^\dagger \phi)^3$. The authors in Ref.~\cite{Grojean:2004xa} provide the preferred region of the scale $\Lambda_6$ of this operator, to facilitate a first-order electroweak phase transition. For a (negative) unity Wilson coefficient of the operator $O_6$, Ref.~\cite{Grojean:2004xa} constraints the scale of this operator $\Lambda_6$ to be,
\beq
\frac {v^4} {m_H^2} < \Lambda_6^2 < \frac {3 v^4} {m_H^2},
\eeq
and the detailed analysis in Ref.~\cite{Delaunay:2007wb} improves the upper limit by about 25\%. The upper bound can be understood from the requirement of the operator $O_6$ being sufficiently sizable to change the Higgs potential to provide a first-order phase transition. 
%Too high a scale of such operator leads to insufficient change in the potential.

By integrating out the singlet field, one can map the general Lagrangian of the singlet extension of the SM to the corresponding SM EFT. The matching is detailed in Ref.~\cite{deBlas:2014mba,Henning:2014wua}, where the EFT operators generated by integrating out the singlet field are explicitly shown for both tree level and one loop level.
For tree-level generation of the $O_6$ operator, the $Z_2$ breaking vertex $s(\phi^\dagger\phi)$ is required. One may anticipate the spontaneous $Z_2$ breaking theory to generate the $O_6$ operator at tree-level as well. However, the two contributing tree level diagrams involving $s^2(\phi^\dagger \phi)$ and $s^3$ cancel each other due to the simple form of the solution to the equation of motion for the singlet field.\footnote{The numerical factors for the two contributions to the $O_6$ operator from tree level diagrams are important. The EFT matching results from the earlier work in Ref.~\cite{Henning:2014gca} without these factors lead to nonvanishing tree level $O_6$ operators in the spontaneous $Z_2$ breaking singlet extension of the SM.}
%This is because integrating out the singlet at tree-level results in the quantum action preserving the $Z_2$ symmetry.
%This is because the tree-level integrating out singlet field results in that the quantum action preserves the classical linear $Z_2$ symmetries in the original Lagrangian, which prohibits the tree-level generation of the $O_6$ operator due to non-existence of $s(\phi^\dagger \phi)$ vertex. 
The Higgs potential is then modified by the singlet field at loop level. Consequently, the scale of the operator is further suppressed by a loop factor of $1/(16\pi^2)$. This results in insufficient modifications to the Higgs potential to trigger a first-order electroweak phase transition. While the EFT is a good description for a one-step phase transition in the electroweak direction, where the singlet field %does not get a vev at zero temperature and 
is heavy enough to be treated as a classical field, the thermal history could be more complex. A detailed study to truly understand the relevant parameter space for sufficiently strong first-order electroweak phase transition is required and we postpone it for future work.

\begin{figure}[t]
  \centering
  \includegraphics[width=1\textwidth]{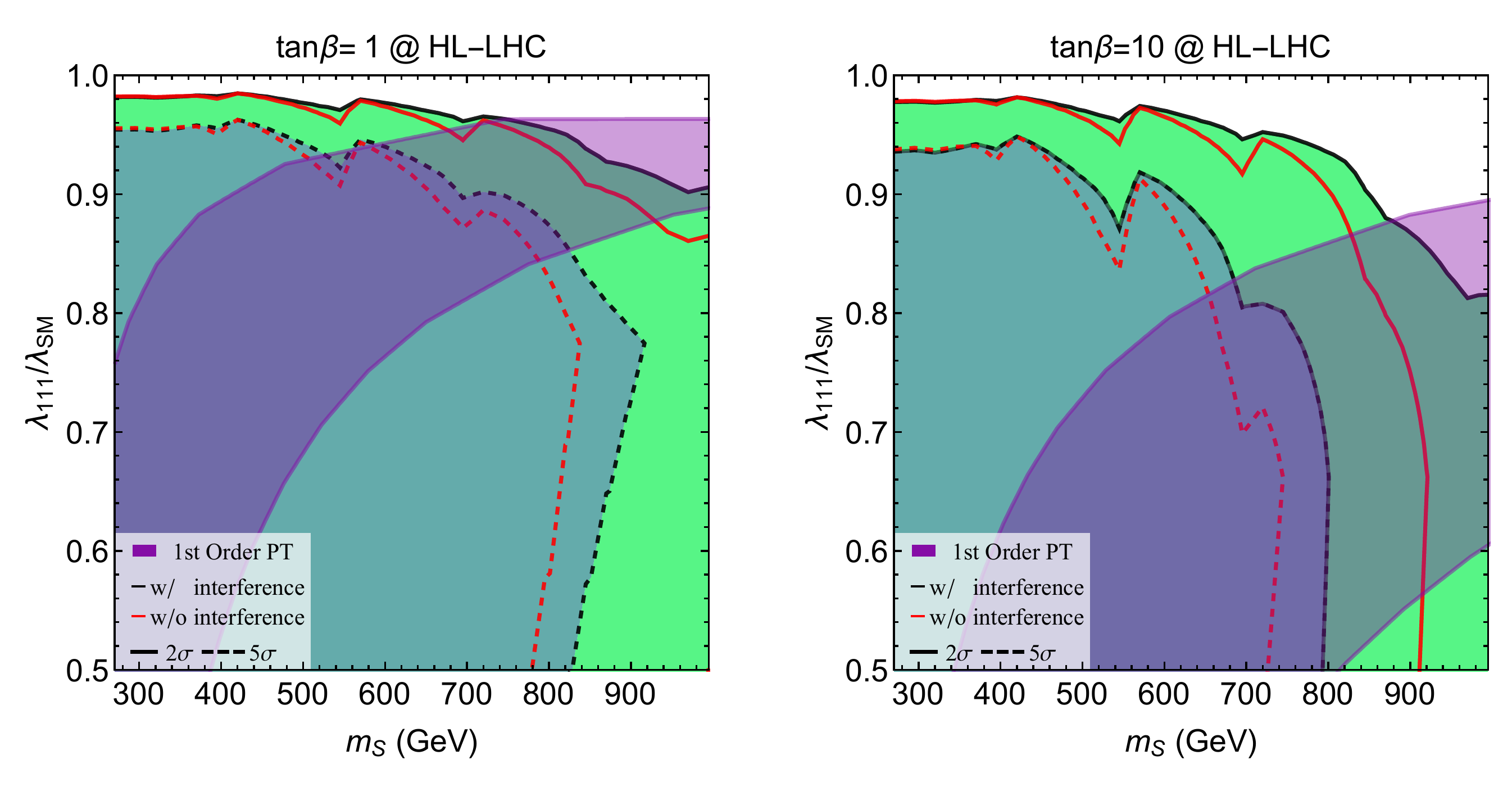}
  \caption{Projected exclusion (solid lines) and discovery (dashed lines) limits at HL-LHC as a function of the heavy singlet scalar mass $m_S$ and the SM-like Higgs trilinear coupling $\lambda_{111}$, normalized to its SM value, for $\tan\beta=1$ (left panel) and $\tan\beta=10$ (right panel), for the explicit $Z_2$-breaking SM plus singlet model scenario. The shaded region within the curves are at the HL-LHC reach. 
  The black and red lines represent the projections with and without the interference effects between the singlet resonance diagram and the SM Higgs pair diagram. 
  The purple shaded areas correspond to parameter regions with a first-order electroweak phase transition from the EFT analysis detailed in the text.
  %The left, middle and right panel corresponds to $\tan\beta$ values of 1, 2 and 10 respectively. The shaded region are disallowed by vacuum stability and perturbative unitarity argument.
  }
  \label{fig:HLPTprojection}
\end{figure}

% and still provide a first order electroweak phase transition. 
To demonstrate the relevance of this interference effect on the first-order electroweak phase transition, we consider, as an example, an explicit $Z_2$ breaking scenario. 
%Still, here we would provide, as an example,given the loop-suppressed modifications to the Higgs potential from the EFT point of view, , we work in an explicit $Z_2$ breaking scenario to demonstrate the relevance to first order electroweak phase transition. 
Without modifying any properties of the phenomenology discussed in this paper (except for the RG running part), we choose the same potential as in Eq.~(\ref{eq:potential}), after the spontaneous symmetry breaking, and flip the sign of the coefficient of the $s^3$ term,\footnote{In the generic, explicit $Z_2$-breaking scenario, $\tan\beta$ is effectively absorbed into the definitions of individual coefficients in the potential. Here, for simplicity, we use the spontaneous $Z_2$ breaking parameterization, and hence keep $\tan\beta$, to avoid a cumbersome redefinition of many of the parameters in the model.}
\beq
-\lambda_s v \tan\beta~s^3 \rightarrow + \lambda_s v \tan\beta~s^3.
\eeq
Hence, one would generate the $O_6$ operator at tree level with 
\beq
\mathcal{L}_{EFT}\supset -\frac {\lambda_{s\phi}^3} {2\lambda_s m_s^2} (\phi^\dagger \phi)^3,
\eeq
where $m_s^2= 2\lambda_s \tan^2\beta~v^2$.
The region preferred by the EFT analysis in this particular explicit $Z_2$ breaking theory requires $\lambda_{s\phi}$ being positive and such condition is also consistent with the EFT potential being bounded from below. This requirement corresponds to positive mixing angle between the singlet and the doublet, $\sin\theta>0$.

Within the above setup, in Fig.~\ref{fig:HLPTprojection} we show the exclusion and discovery projections at the HL-LHC in the singlet mass $m_S$ and Higgs trilinear coupling plane, normalized to the SM Higgs trilinear coupling, $\lambda_{111}/\lambda^{\rm SM}_{111}$. As shown in Sec.~\ref{sec:model}, the Higgs trilinear coupling is modified modestly and the trilinear coupling ratio varies between 0.5 to 1. In the purple band, we show the parameter region consistent with a first-order electroweak phase transition in the EFT analysis.
\footnote{The EFT analysis aims to provide a general picture of the relevance of electroweak phase transition. A detailed thermal history analysis is desirable, especially for singlet masses below 300 GeV.}
 Similar to Fig.~\ref{fig:HLprojection}, we can see that the consistent inclusion of the interference effect improves the reach notably. Most importantly, we observe that the improved discovery and exclusion reach overlaps significantly with the parameter region preferred by the first-order electroweak phase transition.

\section{SUMMARY AND OUTLOOK}
\label{sec:conclude}

In this study, we analyze the interference effects in the $gg\to HH$ process in the presence of a heavy scalar resonance.
We focus on the novel effect of the on-shell interference contribution and discuss it in detail considering the framework of the singlet extension of the SM with spontaneous $Z_2$ breaking. % in the singlet sector.
Such singlet extension of the SM is well motivated as the simplest example compatible with a strong first-order phase transition and consistent with the Higgs boson mass measurements at the LHC.
%a crucial ingredient in explaining the matter-antimatter asymmetry of our universe in current epoch. 

We outline the model setup and relate the model parameters, including quartic and trilinear scalar couplings, to physical parameters such as $m_H$, $v$, $m_S$, $\sin\theta$ and $\tan\beta$.
% in physics parameter basis of heavy scalar mass $m_S$, 
%singlet-doublet mixing angle $\sin\theta$ and the singlet doublet vev ratio $\tan\beta$. 
We find that perturbative unitarity requirements set an upper bound on the mass of the singlet scalar only for low $\tan\beta$, and do not impose significant constraints for moderate values of the mixing angle $\sin\theta$. The heavy scalar total width grows as the third power of its mass, and the decay branching fraction into Higgs pairs varies moderately in the 20\%--40\% range for different regions of the model parameter space.%, shown in \autoref{fig:scalardecay}.

The interference pattern between the resonant heavy scalar contribution and the SM nonresonant triangle and box contributions show interesting features. 
%A detailed decomposition of the interference pattern is presented in \autoref{sec:interference} and 
We highlight the constructive on-shell interference effect that uniquely arises between the heavy scalar resonance diagram and the SM box diagram, due to a large relative phase between the loop functions involved.
% shown in \autoref{fig:phase}. 
%We further quantify the size of this constructive on-shell interference effect by showing its strength in \autoref{fig:interference_central}. 
We observe that the on-shell interference effect can be as large as 40\% of the Breit-Wigner resonance contribution and enhances notably the total signal strength, making it necessary taking into account in heavy singlet searches.

To better evaluate the phenomenological implications of the interference effects in the di-Higgs searches, we carried out a line-shape analysis in the $gg\to HH \to \gamma\gamma b\bar b$ channel, taking into account both the on-shell and off-shell interference contributions. We find that both for the HL-LHC and HE-LHC, the proper inclusion of the interference effects increases the discovery and exclusion reach significantly. %, shown in \autoref{fig:HLprojection} and \autoref{fig:HEprojection}.
Furthermore, using a simplified EFT analysis, we show that the parameter regions where the interference effects are important largely overlap with the regions where a first-order electroweak phase transition is enabled in the singlet extension of the SM. 
%, show in \autoref{fig:HLPTprojection}.

%To summarize, we study the on-shell interference effect in the singlet extension of the SM with a spontaneous $Z_2$ breaking in the singlet sector for the heavy scalar produced through gluon-gluon fusion and decaying into pairs of SM-like Higgs. 
%We find the interference effect could enhanced the resonance production rate up to 40\% in the parameter region considered. Consistently taking both the on-shell and off-shell interference effects between the resonant diagram and the SM Higgs pair diagrams would improve the reach at HL-LHC and HE-LHC significantly. 
%We further discuss the relevance of the parameter regime to first order electroweak phase transition through an EFT analysis in an explicitly $Z_2$ breaking scenario. 
Summarizing, this work shows that a careful understanding of the contributions to the di-Higgs signal in the singlet extension of the SM can be crucial in testing the idea of electroweak baryogensis at colliders.
% in probing the parameter space relevant for matter-antimatter asymmetry
Moreover, our analysis is applicable for a general potential of the singlet extension of the SM as well as for more general resonance searches. A comprehensive analysis of the interference effectsin different decay modes of the Higgs boson and the heavy scalar would provide complementarity information, adding to the LHC potential in the search for heavy scalars.
% should be carried for a complete understanding of the LHC potential for a general heavy scalar searches. 
Furthermore, a detailed study of the electroweak phase transition for the full model with the singlet is an interesting next step that will be presented elsewhere.
%in exploring possibilities for the matter-antimatter asymmetry at the electroweak scale.
%, is an interesting next step to be discussed elsewhere.
\\

{\it Acknowledgment:} We thank Kiel Howe, Ian Lewis, Da Liu, Felix Yu and Yue Zhang for helpful discussion.  
This manuscript has been authored by Fermi Research Alliance, LLC under Contract
No. DE-AC02-07CH11359 with the U.S. Department of Energy, Office of Science, Office of
High Energy Physics. The United States Government retains and the publisher, by accepting
the article for publication, acknowledges that the United States Government retains a nonexclusive,
paid-up, irrevocable, world-wide license to publish or reproduce the published form
of this manuscript, or allow others to do so, for United States Government purposes.
M.R. is supported by la Caixa, Severo Ochoa grant program.

%%%%%%%%%%%%%%%%%%%%%%%%%%%%%%%%%%%%%%%%%%%%%%%%%%%%%%%%%%%%%%%%%%%%%%%%
%%%%%%%%%%%%%%%%%%%% APPENDIX

\appendix
%\newpage
\section{Details on the theoretical constraints}
\label{sec:appthconstraints}

In this section, we show the relevant equations for the stability and perturbative unitarity arguments presented in Sec.~\ref{sec:stabilityunitarityandEWSBconditions}. First, the RGE equations for the quartic couplings of the real singlet model are~\cite{Lerner:2009xg}.
\bea\nonumber
16\pi^2\,&\frac{d}{d\ln\mu} \lambda  &=  24 \lambda^2+\frac{1}{2}\lambda_{s\phi}^2 + 3 \lambda (4y_t^2-3g^2-g^{\prime 2})-6y_t^4+\frac{3}{8}(2g^4+(2g^{\prime 2}+g^2)^2)\\\nonumber
16\pi^2\,&\frac{d}{d\ln\mu}\lambda_s &= 18\lambda_s^2+2\lambda_{s\phi}^2\\
16\pi^2\,&\frac{d}{d\ln\mu}\lambda_{s\phi} &= 4\lambda_{s\phi}^2+6 \lambda_{s\phi}\lambda_s +12\lambda_{s\phi}\lambda + \frac{3}{2} \lambda_{s\phi}(4y_t^2-3g^2-g^{\prime 2}).
\eea
For the analysis we also take into account the running of the top Yukawa $y_t$ and the QCD coupling $g_s$,
\beq
16\pi^2\,\frac{d}{d\ln\mu} y_t  =  \frac{3}{2}y_t^3 - 8 g_s^2y_t -\frac{9}{4}g^2y_t-\frac{17}{12}g^{\prime 2} y_t\,,\quad\quad 
16\pi^2\,\frac{d}{d\ln\mu} g_s = -7g_s^3.
\eeq
Notice that the RGE of $\lambda_{s\phi}$ is proportional to itself, showing the fact that setting it to zero decouples the two sectors. In deriving the limits in Fig.~\ref{fig:thconstraints}, we start the RGE at 1 TeV.

The perturbative unitarity argument (see e.g. Ref.~\cite{Djouadi:2005gi}) is based on the idea that the scattering amplitude can be decomposed into partial waves $a_\ell$ as
\beq
\mathcal{A} = 16\pi \sum_{\ell=0}^{\infty}(2\ell+1)P_\ell(\cos\theta)a_\ell(s).
\eeq
The cross section, proportional to $|a_\ell|^2$, is related to the imaginary part of the forward amplitude through the optical theorem, $\sim \text{Im}(a_\ell)$. This condition, taking the $J=0$ part of the amplitude, corresponds to Eq.~(\ref{eq:pertbound})
\beq
\frac{1}{16\pi s}\int_s^0 dt\,|\mathcal{A}| \,<\, \frac{1}{2}.
\label{eq:a4}
\eeq
To have the theory well defined perturbatively, we require that any $2\to 2$ processes among the vector bosons and scalars satisfy this condition.%, but any combination of initial and final states. 
Therefore we construct the matrix with the scattering amplitudes and require that the largest eigenvalue passes the constraint in Eq.~(\ref{eq:a4}). For completeness we show in the following the leading high energy terms for the different scattering amplitudes of the theory. 

The amplitudes among SM fields are given by,
\bea\nonumber
&\mathcal{A} (WW\to WW) &= -4\lambda -\frac{g^2}{2c_W^2}\frac{s^2+st+t^2}{st}\\\nonumber
&\mathcal{A} (ZZ\to ZZ) &= -6\lambda\\\nonumber
&\mathcal{A} (WW\to ZZ) &= -2\lambda - \frac{g^2}{2} \frac{s^2+st+t^2}{t(s+t)}\\\nonumber
&\mathcal{A} (WW\to HH) &= -2\lambda c_\theta^2-\lambda_{s\phi}s_\theta^2 - \frac{g^2}{2}c_\theta^2 \frac{s^2+st+t^2}{t(s+t)}\,=\, \mathcal{A} (ZZ\to HH) \,[\, g\to g/c_W\,]\\\nonumber
%\mathcal{A} (ZZ\to HH) &=& -2\lambda c_\theta^2-\lambda_{s\phi}s_\theta^2 - \frac{g^2}{2c_W^2}c_\theta^2 \frac{s^2+st+t^2}{t(s+t)}\\\nonumber
&\mathcal{A} (ZH \to ZH) &= -2\lambda c_\theta^2-\lambda_{s\phi}s_\theta^2 - \frac{g^2}{2c_W^2}c_\theta^2 \frac{s^2+st+t^2}{s(s+t)}\\
&\mathcal{A} (WW\to ZH) &= i \frac{g^2}{4c_W^2} c_\theta \left(  1+2\frac{t}{s} +2c_W^2\frac{(s+2t)(s^2+st+t^2)}{st(s+t)} \right)
\eea
These SM amplitudes involving the Higgs boson are with a factor of $c_\theta \equiv \cos\theta$ for each external Higgs due to mixing, and also a contribution proportional to $\lambda_{s\phi}s_\theta$.\footnote{Similarly, $s_\theta\equiv\sin\theta$, $c_{2\theta}\equiv\cos 2\theta$, $c_{4\theta}\equiv\cos 4\theta$, $s_{2\theta}\equiv\sin 2\theta$. $c_W\equiv \cos\theta_W$, where $\theta_W$ is the Weinberg angle.}

The  amplitudes involving only the scalar fields are,
\bea\nonumber
&\mathcal{A} (HH\to HH) &= -6 ( \lambda c_\theta^4 + \lambda_{s\phi}c_\theta^2s_\theta^2 + \lambda_s s_\theta^4 )\\\nonumber
&\mathcal{A} (SS\to SS) &= -6 ( \lambda s_\theta^4 + \lambda_{s\phi}c_\theta^2s_\theta^2 + \lambda_s c_\theta^4 )\\\nonumber
&\mathcal{A} (HH\to SS) &= \mathcal{A} (HS\to HS) = -\frac{3}{4}\left[(\lambda+\lambda_s)(1-c_{4\theta})\,+\, \lambda_{s\phi}(\frac{1}{3}+c_{4\theta})\right]\\\nonumber
&\mathcal{A} (HH\to HS) &= \frac{3}{2}\left[-(\lambda-\lambda_s)-(\lambda+\lambda_s+\lambda_{s\phi})c_{2\theta}\right]s_{2\theta}\\
&\mathcal{A} (SS\to HS) &= \frac{3}{2}\left[-(\lambda-\lambda_s)+(\lambda+\lambda_s-\lambda_{s\phi})c_{2\theta}\right]s_{2\theta}.
\eea

Finally, we show the amplitudes involving SM gauge fields and the singlet $S$,
\bea\nonumber
&\mathcal{A}(WW\to SS) &= -2\lambda s_\theta^2-\lambda_{s\phi}c_\theta^2 - \frac{g^2}{2}s_\theta^2 \frac{s^2+st+t^2}{t(s+t)}\\\nonumber
&\mathcal{A} (ZZ\to SS) &= -2\lambda s_\theta^2-\lambda_{s\phi}c_\theta^2 - \frac{g^2}{2c_W^2}s_\theta^2 \frac{s^2+st+t^2}{t(s+t)}\\\nonumber
&\mathcal{A} (WW\to ZS) &= -i \frac{g^2}{4c_W^2} s_\theta \left(  1+2\frac{t}{s} +2c_W^2\frac{(s+2t)(s^2+st+t^2)}{st(s+t)} \right)\\\nonumber
&\mathcal{A} (ZH\to ZS) &= -2\lambda c_\theta s_\theta + \lambda_{s\phi} c_\theta s_\theta + \frac{g^2}{2c_W^2} c_\theta s_\theta \frac{s^2+st+t^2}{st}\\\nonumber
&\mathcal{A} (ZS\to ZS) &= -2\lambda s_\theta^2 - \lambda_{s\phi} c_\theta^2 +\frac{g^2}{2c_W^2} s_\theta^2 \frac{s^2+st+t^2}{st}\\\nonumber
&\mathcal{A}(WW\to HS) &= -2\lambda c_\theta s_\theta + \lambda_{s\phi} c_\theta s_\theta - \frac{g^2}{2}c_\theta s_\theta \frac{s^2+st+t^2}{t(s+t)}\\
&\mathcal{A}(ZZ\to HS) &= -2\lambda c_\theta s_\theta + \lambda_{s\phi} c_\theta s_\theta - \frac{g^2}{2c_W^2}c_\theta s_\theta \frac{s^2+st+t^2}{t(s+t)}.
\eea

%%%%%%%%%%%%%%%%%%%%%%%%%%%%%%%%%%%%%%%%%%%%%%%%%%%%%%%%%%%%%%%%%%%%%%%%

\section{Indirect constraints}
\label{sec:appthconstraints_indirect}

In this Appendix we summarize the indirect constraints on the singlet model due to Higgs physics precision measurements and EWPO.

\begin{figure}[t]
	\centering 
	\includegraphics[width=1.\textwidth]{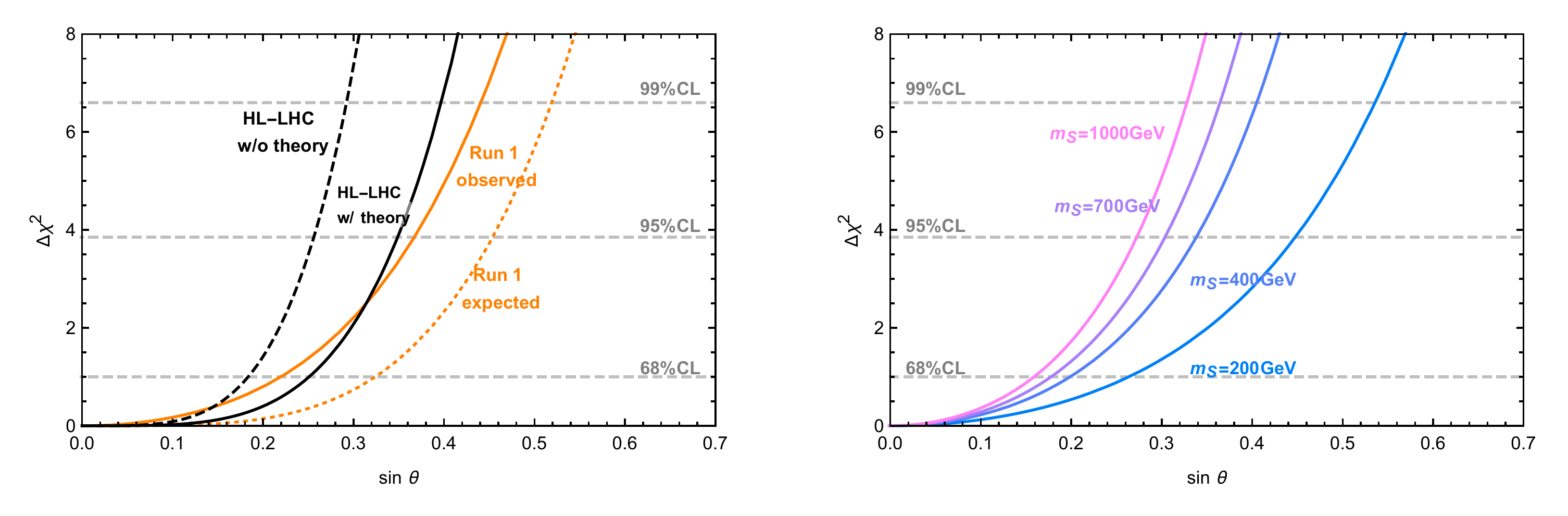}
	\caption{$\Delta\chi^2$ as a function of the mixing angle $\sin\theta$ using different sets of data. \textbf{Left:} Using Higgs signal strengths, with the current LHC constraints and the projected ones at the HL-LHC. \textbf{Right:} Based on LEP constraints EWPO, for different values of the singlet mass.}
	\label{fig:exclusionHiggsmixing}
\end{figure}

The singlet model gives a simple prediction for the signal strengths, since the branching ratios are not modified while the production cross sections are shifted globally by the Higgs mixing. Therefore, the Higgs production is reduced by a factor
\beq
\mu \,=\, 1-\sin^2\theta.
\eeq
The ATLAS and CMS collaborations offered a combination of the LHC runs at 7 and 8 TeV in Ref.~\cite{Khachatryan:2016vau}. The global signal strength $\mu$ of the Higgs production rates, given by %measured to be
\beq
\mu\,=\, 1.09^{+0.11}_{-0.10},
\eeq
is of particular importance since it can be directly applied to constrain the singlet model. The $\chi^2$ analysis for the measured and expected signal is shown in the left panel of Fig.~\ref{fig:exclusionHiggsmixing}. To assess the HL-LHC sensitivity, ATLAS estimates a 3.2\% precision on a global coupling $\kappa$, and the precision can get down to 1.7\% when the theory uncertainties are neglected~\cite{ATL-PHYS-PUB-2014-016}. In our case $\kappa=\sqrt{1-\sin^2\theta}$, and we translate those projections in the figure.

The EWPO from LEP also put constraints on the singlet extension of the SM. %Below the singlet mass, 
%it can be integrated out, generating the operator in the EFT approach.
After integrating out the singlet field, one obtains the EFT operator
%\ZL{$\sin\theta$ not well-defined here; need to re-write as $\lambda_{s\phi}/m_s^2$.}
\beq
\mathcal{L}\supset \, \frac{c_H}{m_s^2}\mathcal{O}_H,%\,=\,\frac{2 \sin^2\theta}{v^2} \,\frac{1}{2} (\partial_\mu \vert H\vert^2)^2
\eeq
where the Wilson coefficient $c_H$ can be matched to the original potential in Eq.~(\ref{eq:potential}) as $c_H=\lambda_{s\phi}^2/(2\lambda_s)$, and the mass scale $m_s$ is approximately the heavy scalar mass $m_S$ in the small mixing angle limit.

The generated EFT operator $O_H$ induces the Higgs field redefinition that shifts the Higgs couplings by $\sim\sin^2\theta$ and also induces other operators through RGE~\cite{Craig:2013xia,Henning:2014gca}. 
In particular, it generates the operator combinations $O_W+O_B$ and $O_T$, which shifts the $S$ and $T$ parameters
\bea%\nonumber
\Delta S &=& +\frac{1} {12\pi} c_H(m_S) \frac {v^2} {m_S^2} \log\left(\frac {m_S^2} {m_W^2}  \right) \\
\Delta T &=& -\frac{3}{16\pi c_W^2} c_H(m_S) \frac {v^2} {m_S^2} \log\left(\frac{m_S^2}{m_W^2}\right).
\eea
Using the electroweak fit in Ref.~\cite{Baak:2012kk,Baak:2014ora} one finds the constraintsopen 
\beq
%S=0.06 \pm 0.09, T = 0.10 \pm 0.07 , \rho = 0.91
S\,=\, 0.06\pm 0.09\,,\quad \quad T\,=\, 0.10\pm 0.07\,,\quad\quad \rho=0.91,
\eeq
where $\rho$ is the correlation coefficient between the $S$ and $T$ parameters.

We show in the right panel of Fig.~\ref{fig:exclusionHiggsmixing} the $\Delta \chi^2$ on the mixing angle for different values of the singlet mass coming from the $S$ and $T$ constraints. We see that the constraints increase with the singlet mass, but a moderate mixing angle of $\sin\theta\sim 0.2$, is still allowed.

\bibliographystyle{JHEP}
\bibliography{diHiggs_SMS_EWPT_submission}

\end{document}